\documentstyle[epsf2,preprint,prd,aps]{revtex}

\tightenlines

\epsfclipon

\def\be{\begin{equation}}
\def\ee{\end{equation}}
\def\ba{\begin{eqnarray}}
\def\ea{\end{eqnarray}}

\def\alt{\ {\raise-3pt\hbox{$\sim$}}\!\!\!\!\!{\raise2pt\hbox{$<$}}\ }
\def\agt{\ {\raise-3pt\hbox{$\sim$}}\!\!\!\!\!{\raise2pt\hbox{$>$}}\ }

\newcommand{\met}{\hbox{{$E_T$}\kern-1.1em\hbox{/}}\kern+0.55em}

\begin{document}

\preprint{OSU-HEP-98-8}

\title {Tau Signals at the Tevatron for Gauge Mediated 
Supersymmetry Breaking}

\author{D. J. Muller and S. Nandi}

\address{Department of Physics, Oklahoma State University,\\
Stillwater, Oklahoma 74078}

\date{November 1998}

\maketitle

\begin{abstract}

We consider the phenomenology of GMSB models where the lighter stau
is the next to lightest supersymmetric particle. In this situation, the
dominant signals for supersymmetry are events with two or three high
$p_T$ $\tau$ leptons accompanied by large missing transverse energy.
We find that the inclusive two $\tau$-jets signature could be observable
at the Tevatron's Run II, while the inclusive three $\tau$-jets
signature could be important at  Run III\@.

\end{abstract}


\section{Introduction}

The phenomenology of gauge mediated supersymmetry breaking (GMSB)
models have been the subject of much interest
lately \cite{{dn},{dmn},{dwt}}. They provide
an alternative to the usually considered case of the soft terms
involving the low energy fields induced by gravity. In GMSB models the
supersymmetry (SUSY) breaking is communicated to the visible
sector by gauge fields. The scale at which this occurs is usually
taken to be around $10^5$\,GeV.

The sparticle spectrum in GMSB models has some significant
differences from the usual gravity mediated SUSY breaking models.
In GMSB models, the gravitino is the lightest supersymmetric
particle (LSP). The next to lightest supersymmetric particle
(NLSP) is typically the lightest neutralino or the lighter stau.
Most phenomenological studies and experimental searches that have
been performed in the context of GMSB have taken the lightest
neutralino to be the NLSP\@. The lightest neutralino then 
decays by $\chi_1^0 \rightarrow \gamma \tilde G$. If this decay
takes place within the detector, the signal involves high $p_T$
photons accompanied by large \met\ \cite{dwr}. For much of the parameter
space, however, the lighter of the two scalar staus is the NLSP\@.
In this case, the decays of SUSY particles produce the $\tilde
\tau_1$ which subsequently decays to a $\tau$ and the gravitino.
If the $\tilde \tau_1$ decays occur within the detector,
signatures for SUSY production will then generally include $\tau$
leptons from the $\tilde \tau_1$ decays and \met\ due to the
stable gravitinos and neutrinos leaving the detector.

It was proposed that GMSB models where the $\tilde\tau_1$ is the
NLSP can lead to unusual and distinguishing signatures for SUSY
production at colliders \cite{nb}.
At the Tevatron, these signals arise from
chargino pair production ($\chi_1^+ \chi_1^-$) and from the
production of the chargino with the next to lightest neutralino
($\chi_1^\pm \chi_2^0$). Slepton pair production ($\tilde l_1^+
\tilde l_1^-$ where $l$ = $e$, $\mu$, or $\tau$) can also be
significant. The subsequent decays typically involve high $p_T$
$\tau$ leptons and substantial \met. In this paper we analyze the
signals for such SUSY production at the Tevatron. In particular we
consider the parameters for the $n = 2$ stau NLSP model line of
the Gauge Mediation/Low Scale SUSY Breaking working group of the
Physics at Run II Supersymmetry/Higgs workshop.

\section{Mass Spectrum and Production Mechanisms}

Since the observed signal depends on the masses of the sparticles, we
first begin by describing the model and the corresponding mass spectrum.
In our model, the messenger sector consists of some number of multiplets
that are
$\bar 5 + 5$ representations of SU(5). They couple to a chiral superfield
$S$ in the hidden sector whose scalar component has a vacuum expectation
value (VEV) $\langle s \rangle$ and whose auxiliary component has a VEV
$\langle F_s \rangle$.
By imposing the requirement that the electroweak (EW) symmetry is broken
radiatively, the
particle spectrum and the mixing angles depend on five parameters:
$M$, $\Lambda$, $n$, $\tan \beta$ and the sign of $\mu$. $M$ is the
messenger scale. $\Lambda$ is equal to
$\langle F_s \rangle$/$\langle s \rangle$ and is related to the SUSY
breaking scale.
The parameter $n$ is dictated by the choice
of the vector-like messenger sector and can take the values 1, 2, 3, or 4
to satisfy the perturbative unification constraint.
The definition of $\tan \beta$ is taken as $\tan \beta \equiv v_2/v_1$ where
$v_2$ is the VEV for the up-type ($H_u$) Higgs doublet and $v_1$ is the
VEV for the down-type ($H_d$) Higgs doublet. The parameter $\mu$ is the
coefficient in the bilinear term, $\mu H_u H_d$, in the superpotential.
Constraints coming from $b \rightarrow s \gamma$
strongly favor negative values for $\mu$ in our convention \cite{ddo}
and, in the cases considered in this work,
$\mu$ is taken to be negative.
Demanding that the EW symmetry be broken radiatively fixes the magnitude
of $\mu$ and the parameter $B$ (from the $B \mu H_u H_d$ term in the
scalar potential) in terms of the other parameters of the theory.

The soft
SUSY breaking gaugino and scalar masses at the messenger scale are
given by \cite{{dn},{spm}}
\be \label{gmass}
\tilde M_i(M) = n \, g(\frac{\Lambda}{M}) \, \frac{\alpha_i(M)}{4 \pi} \,
\Lambda
\ee
and
\be \label{smass}
\tilde m^2(M) = 2\, n \, f(\frac{\Lambda}{M}) \, \sum^3_{i = 1} \, k_i \,
C_i \, \left (\frac{\alpha_i(M)}{4 \pi} \right )^2 \, \Lambda^2
\ee
where the $\alpha_i$ are the three SM gauge couplings and $k_i =$ 1, 1
and 3/5
for SU(3), SU(2), and U(1), respectively. The $C_i$ are zero for gauge
singlets and are 4/3, 3/4 and ($Y$/2)$^2$ for the fundamental
representations of SU(3), SU(2) and U(1), respectively (with $Y$ given
by $Q = I_3 + Y/2$). $g(x)$ and $f(x)$ are messenger scale threshold
functions. We calculate the sparticle masses at the scale $M$
using Eqs.~(\ref{gmass}) and (\ref{smass}) and run these
to the electroweak scale using the appropriate renormalization group
equations \cite{bbo}.

The decay chains and hence the signatures for the events depend on the
particles initially produced as well as the hierarchy of the masses.
Since SUSY breaking is communicated to the visible sector by gauge
interactions, the mass differences between the superparticles depend
on the their gauge interactions. This creates a hierarchy in mass between
electroweak and strongly interacting sparticles. Eq.\,(\ref{gmass}) shows
that the gluino is more massive than the EW charginos and neutralinos,
while Eq.\,(\ref{smass}) shows that squarks are considerably more
massive than sleptons.
Given this hierarchy of sparticle masses and the current lower bounds on
squark and gluino masses, the production
of strongly interacting sparticles is probably not a viable search modes
for SUSY at the Tevatron Run II\@.
A more likely mechanism for producing SUSY particles is
via EW gaugino production. At the Tevatron, chargino pair
($\chi_1^+ \chi_1^-$) production takes place through s-channel $Z$ and
$\gamma$ exchange and $\chi_2^0 \, \chi_1^\pm$ production is through
s-channel
$W$ exchange. Squark exchange via the t-channel also contributes to both
processes, but the contributions are expected to be negligible since the
squark masses are large in GMSB models. The production of
$\chi^0_1 \, \chi^{\pm}_1$ is suppressed due to the smallness of the coupling
involved.
In regions of the parameter space where the production of charginos and
neutralinos is kinematically suppressed, the pair production of sleptons
($\bar{\tilde{\tau}}_1 \tilde{\tau}_1$, $\bar{\tilde{\mu}}_1 \tilde{\mu}_1$
and $\bar{\tilde{e}}_1 \tilde{e}_1$) can be important. Their production
occurs through s-channel $Z$ and $\gamma$ exchange.

Given the hierarchy of sparticle masses in GMSB models, there are roughly
four possible cases to consider for SUSY production at the Tevatron:
\begin{description}
\centering
\item[Case 1:] $m_{\tilde \nu} > M_{\chi_2^0} \agt M_{\chi_1^\pm}
> m_{\tilde e_1, \tilde \mu_1} > M_{\chi_1^0} > m_{\tilde \tau_1}$
\item[Case 2:] $M_{\chi_2^0} \agt M_{\chi_1^\pm} > m_{\tilde \nu}
> M_{\chi_1^0} > m_{\tilde e_1, \tilde \mu_1} > m_{\tilde \tau_1}$
\item[Case 3:] $M_{\chi_2^0} \agt M_{\chi_1^\pm} > m_{\tilde \nu}
> m_{\tilde e_1, \tilde \mu_1} > M_{\chi_1^0} > m_{\tilde \tau_1}$
\item[Case 4:] $m_{\tilde \nu} > M_{\chi_2^0} \agt M_{\chi_1^\pm}
> M_{\chi_1^0} > m_{\tilde e_1, \tilde \mu_1} > m_{\tilde \tau_1}$
\end{description}
The three sneutrino masses are nearly the same. The lighter of the selectrons
and smuons are essentially right handed and have the same mass. Also, for all
the parameter points we considered, $\chi^{\pm}_1$ and $\chi^0_2$
are nearly degenerate in mass.

The possible final state configurations at the Tevatron depend on the
sparticle spectrum, but they will have certain aspects in common. Since
the $\tilde\tau_1$ is the NLSP,
the various possible decay modes will (usually) produce at least
two $\tau$ leptons arising from the decays of the $\tilde\tau_1$'s.
In addition, there can also be large \met\ due to the
stable gravitinos and neutrinos escaping detection.

A special situation in cases 2 and 4 arises when the lighter selectron and the
lighter smuon are nearly degenerate in mass to the lighter stau. When the
mass difference between the slectron and the stau is less than twice
the mass of the $\tau$, essentially the only decay mode for the selectron
is $\tilde{e} \rightarrow e \, \tilde{G}$.
The lighter smuon likewise decays via
$\tilde{\mu} \rightarrow \mu \, \tilde{G}$.
This situation is refered to as the ``co-NLSP" case \cite{akm}.

\section{Analysis and Results}

We now give a detailed analysis of the possible Tevatron signatures for
SUSY production in the context of GMSB models where the lighter stau
is the NLSP and decays promptly within the detector.
This analysis is performed in the context of the Main Injector (MI)
and TeV33 upgrades of the Tevatron collider. The center of mass energy
is taken to be $\sqrt{s} = 2$\,TeV and the integrated luminosity is
taken to be 2\,fb$^{-1}$ for the MI upgrade and 30\,fb$^{-1}$ for the
TeV33 upgrade \cite{TeV33}.

In performing this analysis,
the cuts employed are that final state charged leptons must have
$p_T > 10$\,GeV and a
pseudorapidity, $\eta \equiv -\ln ( \tan \frac{\theta}{2} )$ where
$\theta$ is the polar angle with respect to the proton beam direction,
of magnitude less than 1.
Jets must have $E_T > 10$\,GeV and $|\eta| < 2$.
In addition, hadronic final states within a cone
size of $\Delta R \equiv \sqrt{ (\Delta \phi)^2 + (\Delta \eta)^2 } = 0.4$
are merged to a single jet. Leptons within this cone radius of a jet are
discounted.
For a $\tau$-jet to be counted as such, it must have $|\eta| < 1$.
The most energetic $\tau$ jet is required to have $E_T > 20$\,GeV\@.
In addition, a missing transverse energy cut of \met\ $> 30$\,GeV is imposed.

We consider each mass case in turn. In our analysis, we restrict
ourselves to those regions of the parameter space where the
$\tilde\tau_1$ decays promptly to a $\tau$ and a gravitino. The
parameter space is also restricted to those regions where
$m_{\tilde\tau_1} \agt 70$ GeV\@. Results from LEP-2 set a lower
bound of $m_{\tilde{\tau}_1} \geq 72$\,GeV \cite{pc}.

\section{Stau NLSP line with $n = 2$}

\begin{figure}
\centering
\epsfxsize=0.98\textwidth
\epsfbox[66 143 491 656]{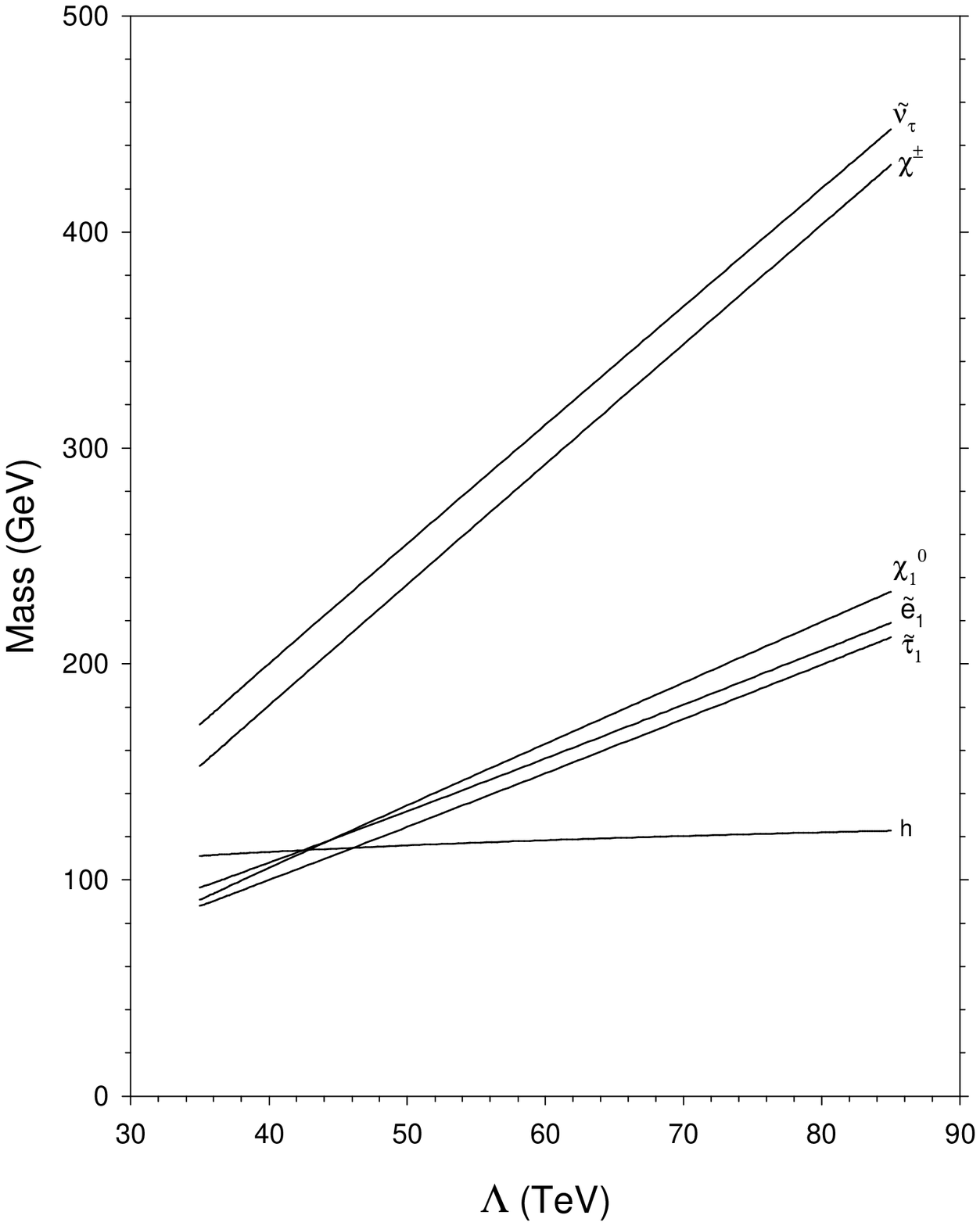}
\vskip 0.25cm
\caption{\label{mass1} Masses for the sparticles of interest for the line
defined by $n = 2$, $\tan \beta = 15$ and $M$/$\Lambda$ = 3.
$M_{\chi_2^0} \approx M_{\chi_1^\pm}$ and
$M_{\tilde{\mu}_1} \approx M_{\tilde{e}_1}$.  }
\end{figure}

In this section we do the analysis for points along the line defined
by the parameter values $M$/$\Lambda$ = 3, $n = 2$ and $\tan \beta = 15$.
We vary $\Lambda$ from 35\,TeV to 85\,TeV\@. The masses for the sparticles
that are of interest here are given in Fig.~\ref{mass1}. Note that the
sneutrino mass is always above that of the lightest chargino and the
lightest two neutralinos. Thus the sneutrinos do not figure into the
decay chains of the major SUSY production mechanisms here. Note that the
lightest neutralino is below the selectron/smuon mass at the lower
end of the $\Lambda$ scale ($\Lambda \alt 45$\,TeV).
Thus for $\Lambda \alt 45$\,TeV the mass spectrum is of type 1.
For $\Lambda \agt 45$\,TeV, the mass spectrum is of type 4. In this
region of $\Lambda$, there are more decay modes for the various particles
due to the increasing masses of all the sparticles as well as the shift
in the position of the lightest neutralino in the mass hierarchy.

\begin{figure}
\centering
\epsfxsize=0.98\textwidth
\epsfbox[59 151 494 664]{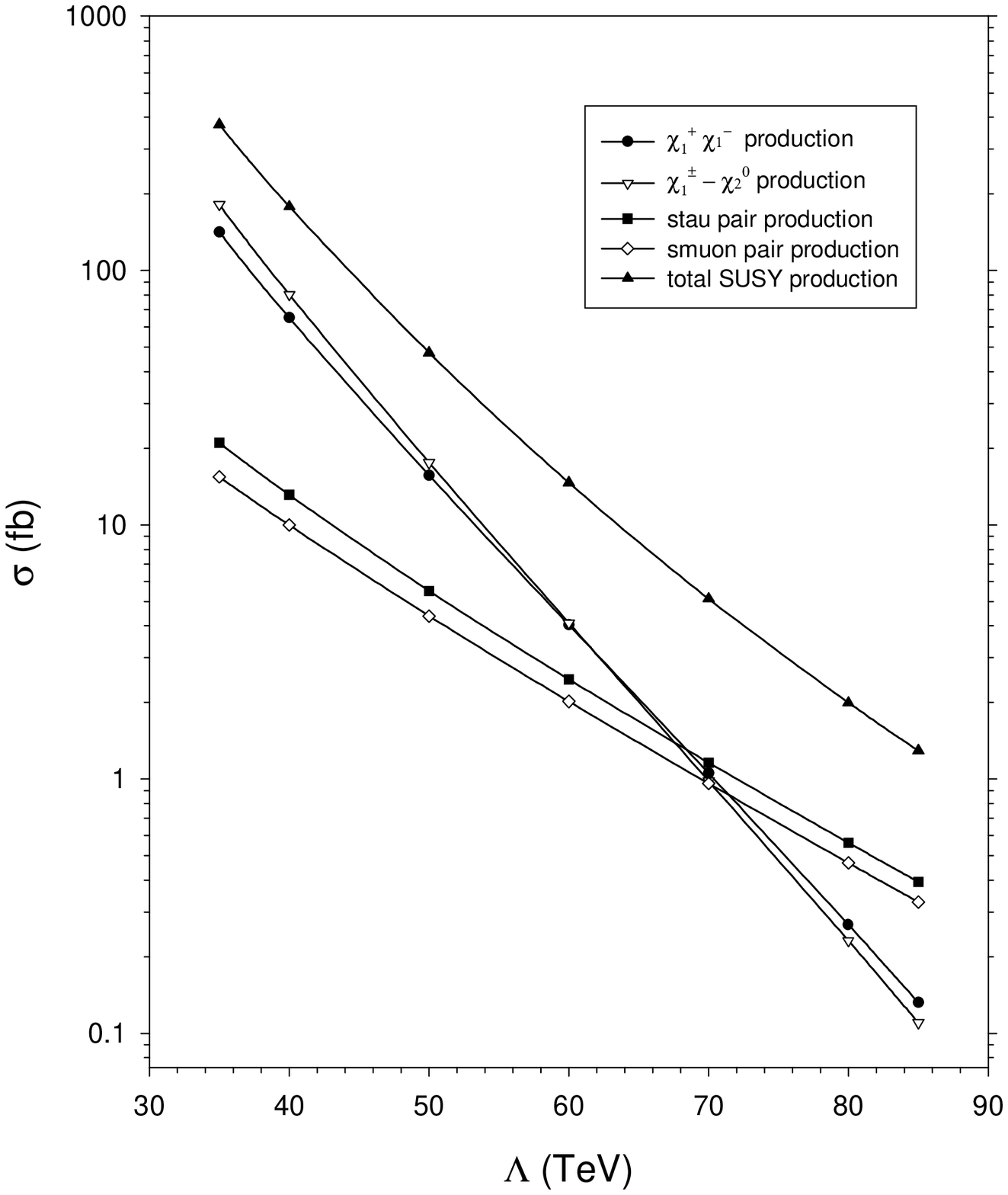}
\vskip 0.25cm
\caption{\label{cross1} Cross section for the important SUSY production
processes at the Tevatron for the line defined by $n = 2$,
$\tan \beta = 15$ and $M$/$\Lambda$ = 3. The $\chi_2^0 \chi_1^\pm$ cross
section includes production of both signs of the chargino.}
\end{figure}

The cross sections for these parameters are given in Fig.~\ref{cross1}.
From the figure, the cross sections for $\chi_1^+ \chi_1^-$ and
$\chi_2^0 \chi_1^\pm$ production dominate for the region where
$\Lambda$ is below 65\,TeV\@.
As $\Lambda$ increases, the masses of the gauginos increase
significantly and hence the cross section falls off.
For $\Lambda \agt 70$\,TeV, the production rates for
$\tilde{\tau}_1^+ \tilde{\tau}_1^-$ and $\tilde{e}_1^+ \tilde{e}_1^-$
are dominant, but the cross sections tend to be rather low.

The signature depends on the allowed decay modes of the sparticles and
their branching ratios. The branching ratios for the sparticles of
interest are given in Table~\ref{br1}. Since $\tilde{\tau}_1$ is
the NLSP, it decays via $\tilde{\tau}_1 \rightarrow \tau \, \tilde{G}$.
The decays of the selectron and smuon depend strongly on the value
of $\Lambda$. For $\Lambda$ below $\sim 45$\,TeV, the lightest
neutralino has a mass below that of $\tilde{\mu}_1$ and $\tilde{e}_1$.
As a consequence of this, the main decay mode of lightest smuon is
$\tilde{\mu}_1 \rightarrow \chi_1^0 \, \mu$ and the main decay
mode of the lightest selectron is $\tilde{e}_1 \rightarrow \chi_1^0 \, e$.
For higher values of $\Lambda$, however, the lightest neutralino mass
increases above that of $\tilde{e}_1$ and $\tilde{\mu}_1$. Then the
only available two-body decay mode for the lightest smuon is
$\tilde{\mu}_1 \rightarrow \mu \, \tilde{G}$ with the corresponding
decay mode for the lightest selectron. Given the smallness of the
coupling involved, though, the possibility exists that some three-body
decays could be important.
Indeed, the neutralino mediated decays
$\tilde{\mu}_1^- \rightarrow e^- \tau^- \tilde{\tau}^+$ and
$\tilde{\mu}_1^- \rightarrow e^- \tau^+ \tilde{\tau}^-$ are the important
decay modes \cite{akm2} for these higher values of $\Lambda$.

Since the lightest neutralino
tends to be one of the least massive sparticles, its only decay modes
are $\chi_1^0 \rightarrow \tilde{\tau}_1 \tau$ and the decays to
$\tilde{\mu}_1$ and $\tilde{e}_1$ if $\chi_1^0$ is greater in mass
than those sparticles. Since the lightest neutralino is less
massive than the selectrons and smuons for $\Lambda$, the only decay
mode is $\chi_1^0 \rightarrow \tilde{\tau}_1 \tau$. As $\Lambda$
increases above 45\,TeV, $\tilde{\mu}_1$ and $\tilde{e}_1$ become
increasingly important, although
$\chi_1^0 \rightarrow \tilde{\tau}_1 \tau$ remains the dominant decay
mode.

Since the lightest chargino is
mostly wino, it couples mainly to ``left-handed" sfermions. Thus the
decay mode $\chi_1^\pm \rightarrow \tilde{\tau}_1 \nu_\tau$ is
typically important due to the significant mixing of the left-handed and
right-handed staus and the lower mass of the $\tilde{\tau}_1$.
This decay mode is, in fact, essentially the only decay mode for low
values of $\Lambda$ with the parameters considered here.
Thus with the subsequent decay
$\tilde{\tau}_1 \rightarrow \tau \, \tilde{G}$, there are typically
two $\tau$ leptons produced in $\chi_1^+ \chi_1^-$ production at
these values of $\Lambda$.
As $\Lambda$
increases, however, the decay mode
$\chi_1^\pm \rightarrow \chi_1^0 W$ becomes available and becomes the
dominant decay mode as $\Lambda$ increases above 60\,TeV\@. With the
two $\tau$ leptons that can be expected from the lightest neutralino
decay and the $W \rightarrow \tau \nu_\tau$ decay, we can expect
up to six $\tau$ leptons from $\chi_1^+ \chi_1^-$ production at these
larger values of $\Lambda$.

There are many decay modes for the second lightest neutralino as
Table~\ref{br1} shows. For low values of $\Lambda$, the dominant decay
mode is $\chi_2^0 \rightarrow \tilde{\tau}_1 \tau$ at 50 - 60\%.
The decays to the
other sleptons are also important at 15 - 20\%.
Thus $\chi_1^\pm \chi_2^0$ production produces three $\tau$ leptons:
two from
the slepton decays of the neutralino and one from the decay
$\chi_1^\pm \rightarrow \tilde{\tau}_1 \, \nu_\tau$ followed by
$\tilde{\tau}_1 \rightarrow \tau \, \tilde{G}$.
As $\Lambda$ increases above 55\,TeV,
The decay $\chi_2^0 \rightarrow \chi_1^0 {\rm h}$, where h is the
Higgs boson,
rapidly becomes the dominant decay mode. The decay
$\chi_2^0 \rightarrow \chi_1^0 \, Z$ is also present,
but of relatively little importance.

\begin{figure}
\centering
\epsfxsize=0.6\textwidth
\epsfbox[84 233 487 544]{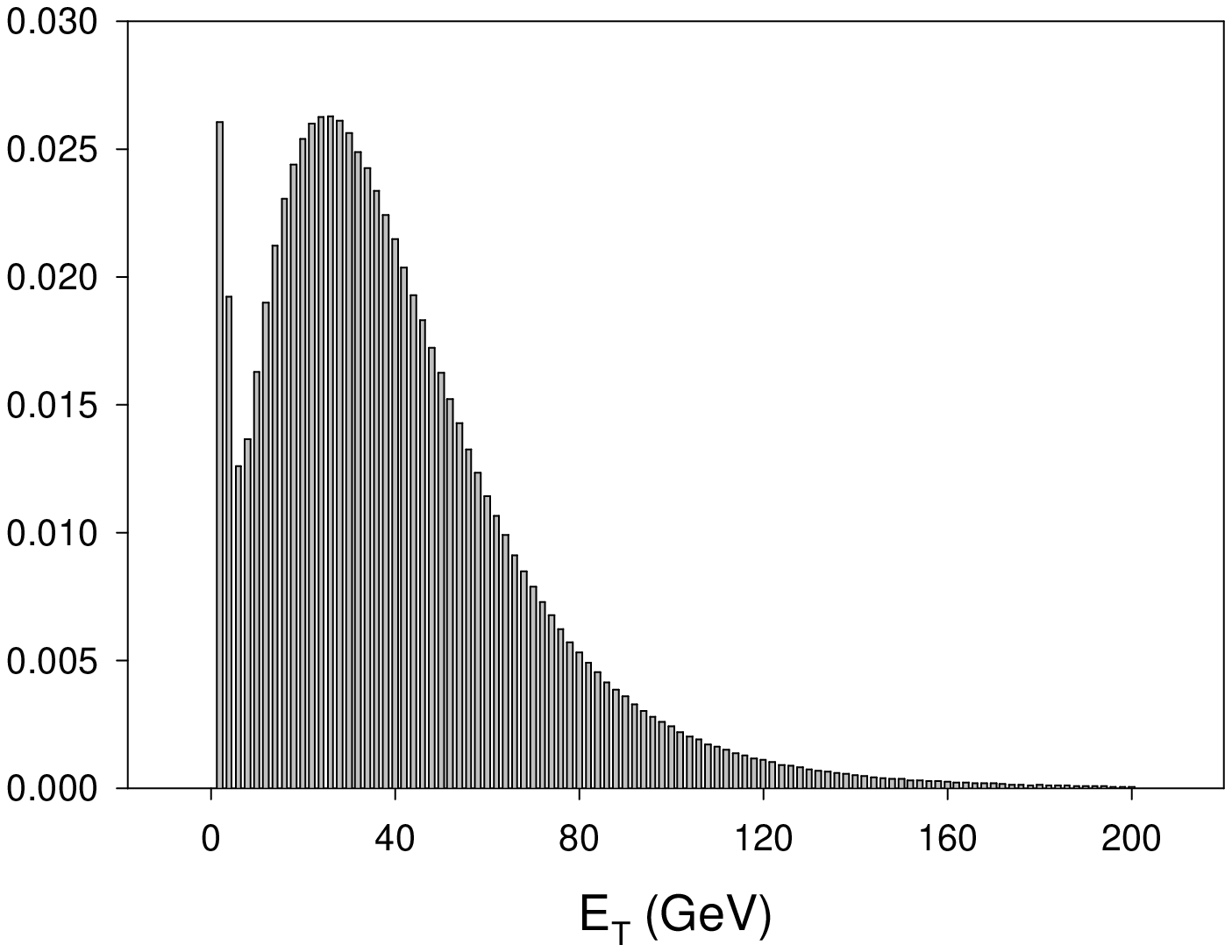}
\begin{minipage}{0.6\textwidth}
(b)
\end{minipage}
\vspace{1.5cm}
\caption{\label{let35} The $E_T$ distributions of the leading $\tau$ jet
for the parameters $n = 2$, $\tan\beta = 15$, $M$/$\Lambda$ = 3 and
$\Lambda = 35$\,TeV\@.
In (a), no cuts have been imposed.
In (b), the  $|\eta| < 1$ cut has been imposed.}
\end{figure}

\begin{figure}
\centering
\epsfxsize=0.6\textwidth
\epsfbox[84 233 487 544]{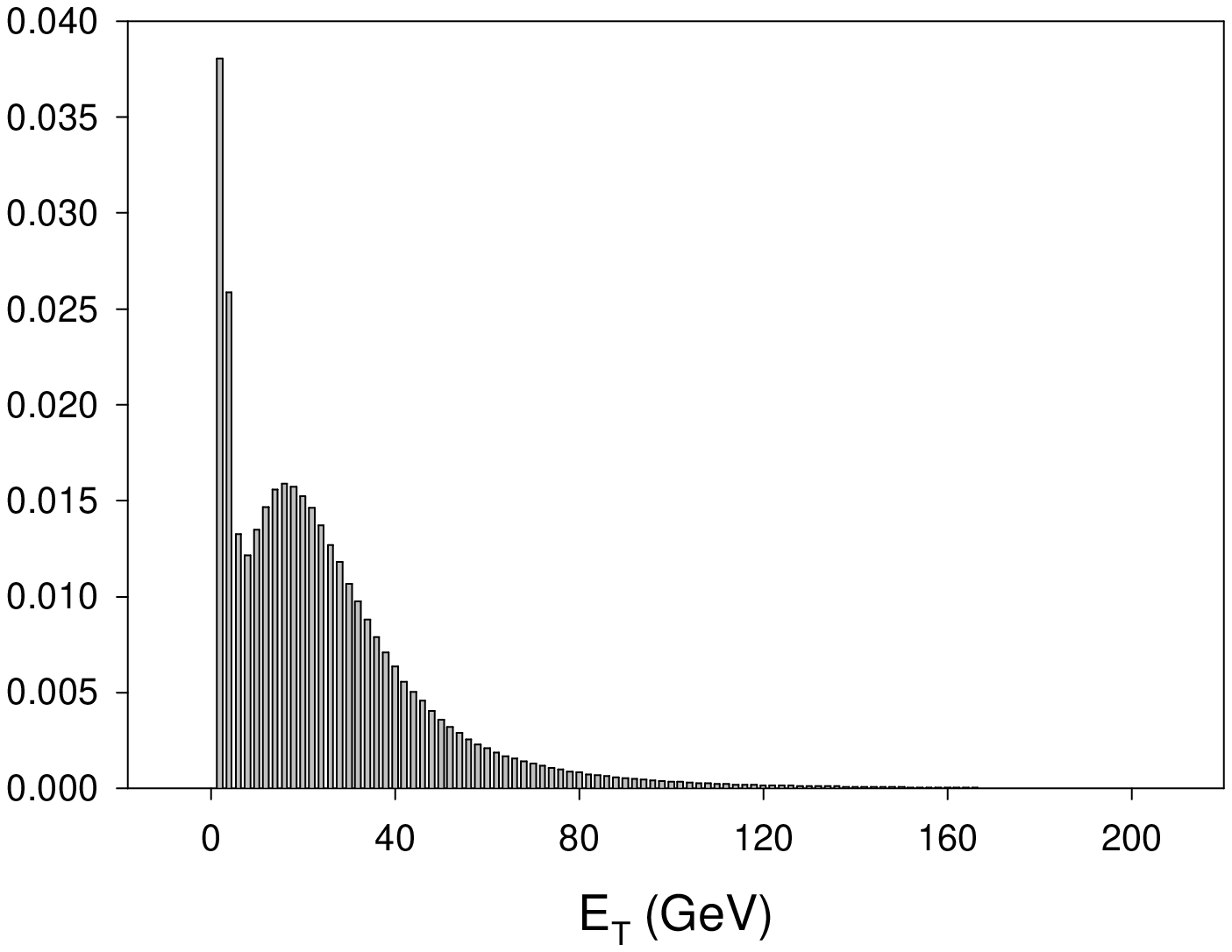}
\begin{minipage}{0.6\textwidth}
(b)
\end{minipage}
\vspace{1.5 cm}
\caption{\label{set35} The $E_T$ distributions of the secondary $\tau$ jet
for the parameters $n = 2$, $\tan\beta = 15$, $M$/$\Lambda$ = 3 and
$\Lambda = 35$\,TeV\@.
In (a), no cuts have been imposed.
In (b), the  $|\eta| < 1$ cut on $\tau$-jets has been imposed.}
\end{figure}

Given the cuts that we place on the $\tau$-jets, the question
arises as to how high we can expect the $E_T$ of the $\tau$-jets to be.
Fig.~\ref{let35} gives the $E_T$ distribution of the highest $E_T$
$\tau$-jet for $\Lambda = 35$\,TeV\@. The pseudorapidity cut of
$|\eta| < 1$ on $\tau$-jets has been imposed in Fig.~\ref{let35}(b).
The peak in the distribution occurs at about 25\,GeV with a broad tail
that reaches out beyond 120\,GeV\@. Thus the leading $\tau$-jets are
relatively hard and many will pass the transverse energy cut of
$E_T > 20$\,GeV\@. The next to highest $E_T$ $\tau$-jet is
significantly different as Fig.~\ref{set35} shows. Here the distribution
peaks at a lower value of about 15\,GeV and hardly extends at all beyond
80\,GeV\@. Due to the softness of the secondary $\tau$-jets, many of the
$\tau$-jets will tend to be eliminated by the cuts.

\begin{figure}
\centering
\epsfxsize=0.6\textwidth
\epsfbox[84 233 487 544]{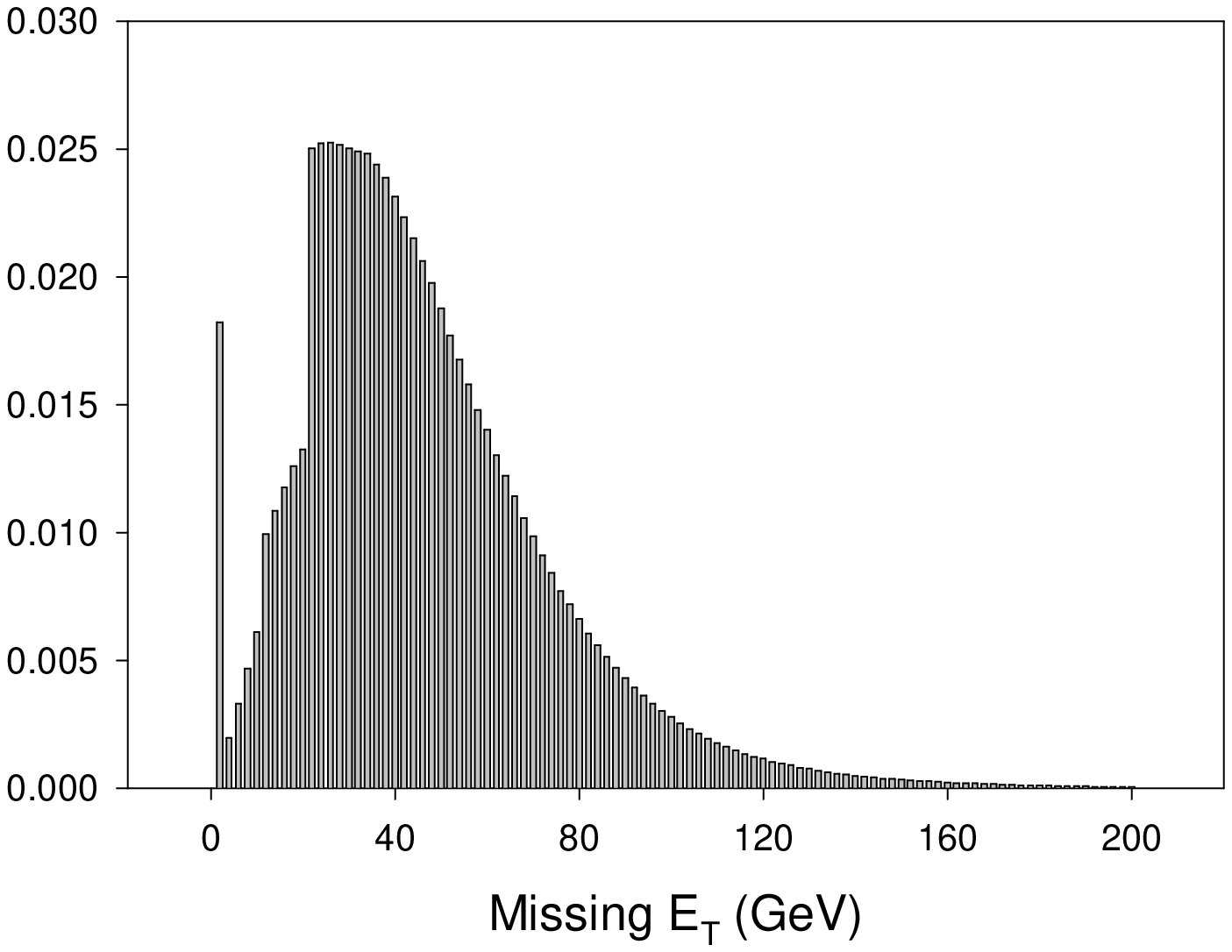}
\begin{minipage}{0.6\textwidth}
(b)
\end{minipage}
\vspace{1.5 cm}
\caption{\label{met35} \met\ distribution of the secondary $\tau$ jet for
the parameters $n = 2$, $\tan\beta = 15$, $M$/$\Lambda$ = 3 and
$\Lambda = 35$\,TeV\@.
In (a), no cuts have been imposed.
In (b), the $E_T$/$p_T$ and pseudorapidity cuts on the jets and charged
leptons have been imposed.}
\end{figure}

Also of interest is the \met\ distribution. With energetic and stable
gravitinos and neutrinos produced in the decays, it is expected that
large missing transverse energy could be an important part of the
signal. Since the missing transverse energy is calculated from what
is observed, however, the question arises as to whether significant
cancellation occurs due to the many decay products. Fig.~\ref{met35}
gives the \met\ distribution for the case where $\Lambda = 35$\,TeV\@.
The figure demonstrates that the \met\ distribution is indeed broad
with a tail reaching out beyond 120\,GeV\@. The peak before cuts
occurs at about 35\,GeV and the peak still occurs at about 35\,GeV when
$E_T$/$p_T$ and pseudorapidity cuts are applied to the various particles.
Thus a 30\,GeV cut should not be too restrictive. As $\Lambda$ is
increased, the \met\ distribution gets harder since the gaugino masses
get larger as $\Lambda$ is increased.

We now consider the specifics of the various final state
possibilities. Table~\ref{tau1br35} give the inclusive branching
ratios for different number of $\tau$-jets for $\Lambda =
35$\,TeV\@. As indicated above, this example always produces two
$\tau$ leptons in chargino pair production. Before cuts the
inclusive branching ratio for the 2 $\tau$-jet mode in chargino
pair production is 42\%, while the 1 $\tau$-jet mode in chargino
pair production is 45.6\%. After the cuts specified above, the
branching ratios are cut down rather substantially. The one
$\tau$-jet BR becomes 25.8\% and the two $\tau$-jet BR is 10.8\%.
The situation changes as $\Lambda$ increases. This is demonstrated
in Tables~\ref{tau1br50} and \ref{tau1br70} which are for $\Lambda
= 50$ and 70\,TeV, respectively. We see that there is now the
possibility for many more $\tau$-jets. This is due to the
appearance of the decay mode $\chi_1^\pm \rightarrow \chi_1^0 \,
W$. With the decay $\chi_1^0 \rightarrow \tilde{\tau}_1 \tau$
followed by $\tilde{\tau}_1 \rightarrow \tau \, \tilde{G}$ along
with the decay $W \rightarrow \tau \nu_\tau$, there is now the
possibility for producing up to six $\tau$-jets. At $\Lambda =
50$\,TeV, the BR for three $\tau$-jets is 16.3\%. This is cut down
substantially after cuts, however, and the BR becomes only 1\%.
$\Lambda = 70$\,TeV shows similar results, although at this point
this point the cross section for $\chi_1^+ \chi_1^-$ production is
low.

Considering $\chi_2^0 \, \chi_1^\pm$ production, we recall that at
$\Lambda \alt 45$\,TeV, the decay modes of $\chi_2^0$ are
$\chi_2^0 \rightarrow \tilde{\tau}_1 \tau$, $\chi_2^0 \rightarrow
\tilde{\mu}_1 \mu$ and $\chi_2^0 \rightarrow \tilde{e}_1 e$. With
the subsequent decays $\tilde{\tau}_1 \rightarrow \tau \,
\tilde{G}$ and $\tilde{\mu}_1 \rightarrow \mu \tau \tilde{\tau}_1$
along with the corresponding decay of the selectron, the number of
$\tau$ leptons from $\chi_2^0$ decay is two. With the one $\tau$
lepton from the chargino, we have up to three $\tau$ leptons from
$\chi_2^0 \chi_1^\pm$ production at these values of $\Lambda$. We
see from Table~\ref{tau1br35} that for $\Lambda = 35$\,TeV, the
branching ratios for inclusive production of 3 $\tau$-jets is
27.2\% before cuts, while for 1 and 2 $\tau$-jets the branching
ratios are 24\% and 44.3\%, repsectively. These branching ratios
are cut back considerably once the cuts are included. The 3
$\tau$-jet BR in particular is reduced to only 2.6\%. As $\Lambda$
increases, the potential exists to create more than three
$\tau$-jets due to the new decay modes for $\chi_1^\pm$ and
$\chi_2^0$. The branching ratios for more than three $\tau$-jets
tend to be small after cuts, however, as Tables~\ref{tau1br50} and
\ref{tau1br70} demonstrate.

Slepton production tends to be rather simple since there are
relatively few decay modes available to the sleptons. This is
especially true for the production of the lighter stau as its only
significant decay mode is $\tilde{\tau}_1 \rightarrow \tau \,
\tilde{G}$. Thus for $\tilde{\tau}_1^+ \tilde{\tau}_1^-$
production up to two $\tau$-jets are possible. Before cuts, the 2
$\tau$-jet BR is 42\% and the 1 $\tau$-jet BR is 45.6\% for
$\tilde{\tau}_1^+ \tilde{\tau}_1^-$ production independent of the
values of $\Lambda$ considered. After cuts, these drop down to 10
- 17\% and 30 - 40\%, respectively, as shown in
Tables~\ref{tau1br35}, \ref{tau1br50} and \ref{tau1br70}. The
three-body decay modes of the lightest selectron and smuon mean
that up to four $\tau$-jets are possible in $\tilde{\mu}_1^+
\tilde{\mu}_1^-$ production and $\tilde{e}_1^+ \tilde{e}_1^-$
production. The branching ratios for three and four $\tau$-jets in
$\tilde{e}_1^+ \tilde{e}_1^-$ and $\tilde{\mu}_1^+
\tilde{\mu}_1^-$ production are greatly diminished after cuts.

\begin{figure}
\centering
\epsfxsize=0.98\textwidth
\epsfbox[62 124 488 631]{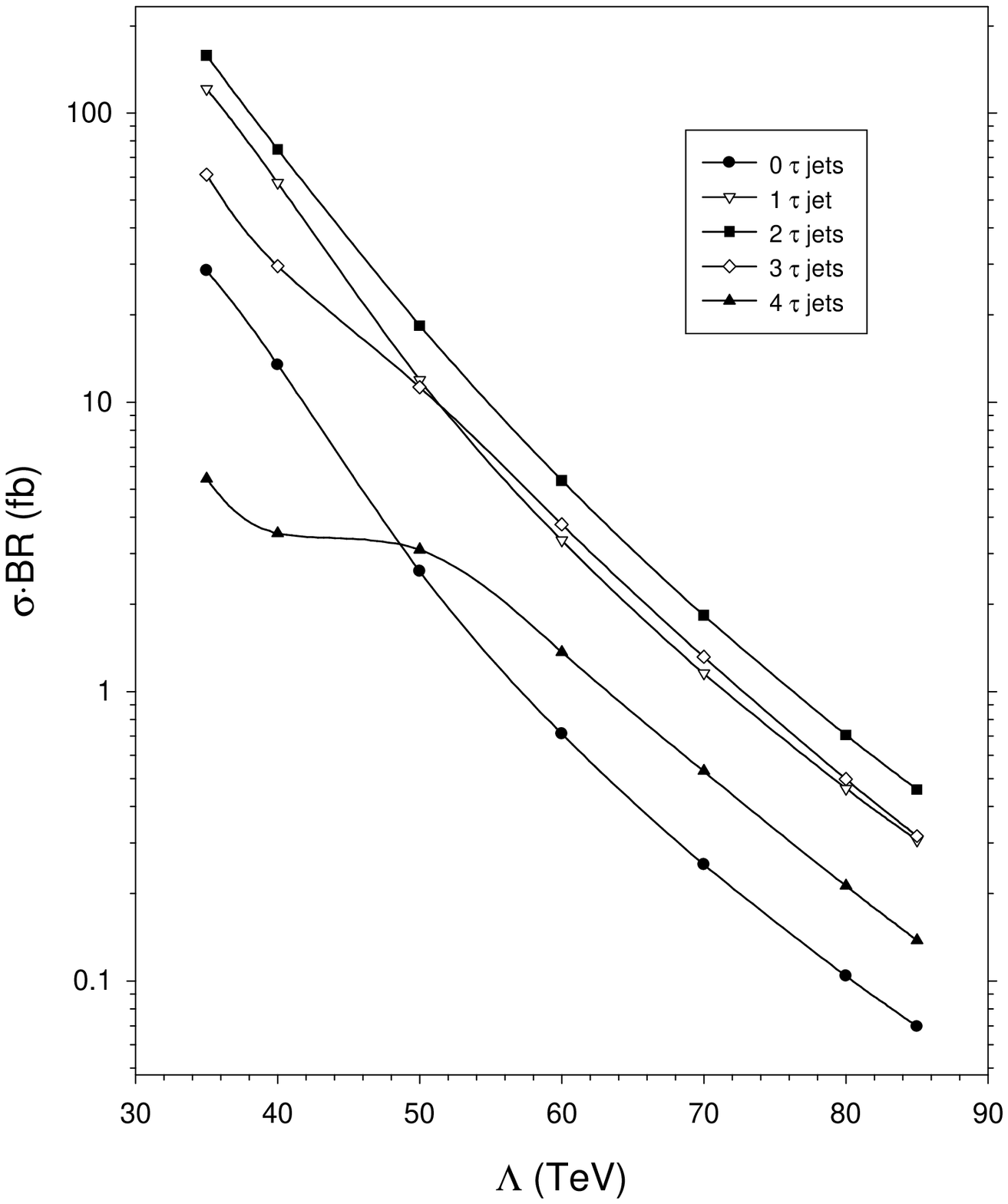}
\vskip 0.25cm
\caption{\label{tauj1nc} $\sigma \cdot BR$ before cuts for the inclusive
$\tau$ jets modes
for the parameters $n = 2$, $\tan \beta = 15$ and $M$/$\Lambda$ = 3.}
\end{figure}

The question now arises as to the observability of these modes at the
Tevatron's Run II\@. The cross sections for inclusive $\tau$-jet
production before cuts are given in Fig.~\ref{tauj1nc}. All the
SUSY production modes considered in this analysis are included.
Events with more than four $\tau$-jets are not included
in the figure due to their
extremely low branching ratios. By far the dominant decay mode is the
2 $\tau$-jets mode. The 3 $\tau$-jets and 1 $\tau$-jet modes are also
quite large.

\begin{figure}
\centering
\epsfxsize=0.98\textwidth
\epsfbox[89 120 515 634]{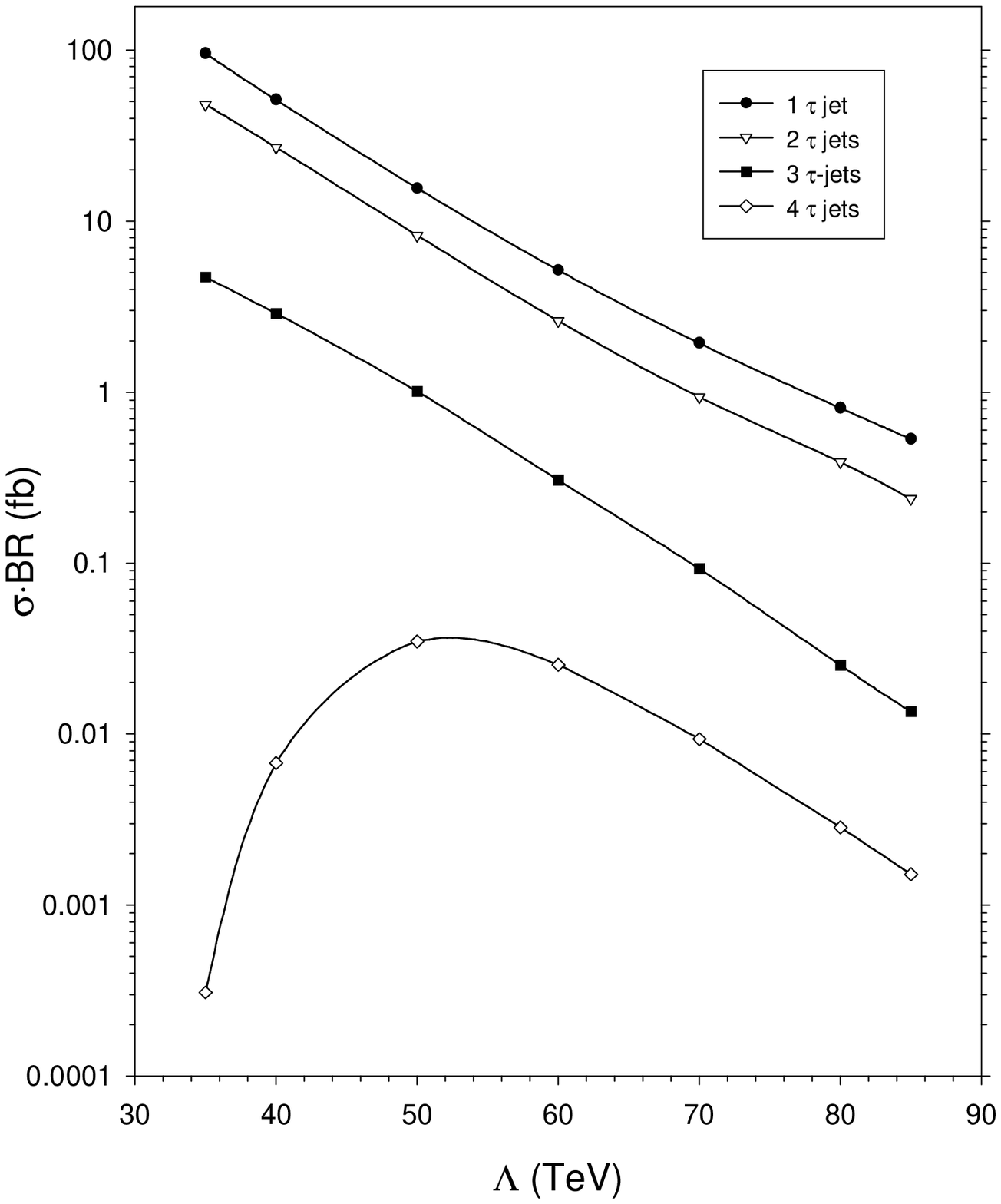}
\vskip 0.25cm
\caption{\label{tauj1c} $\sigma \cdot BR$ after cuts for the inclusive
$\tau$ jets modes
for the parameters $n = 2$, $\tan \beta = 15$ and $M$/$\Lambda$ = 3.}
\end{figure}

Of course, the real issue is what the production cross sections
are after the cuts have been imposed. These are given in
Fig.~\ref{tauj1c}. The graph shows that after cuts, the 1
$\tau$-jet mode is dominant. The 2 $\tau$-jets mode is of the same
order of magnitude and the 3 $\tau$-jet mode is not unappreciable.
For $\Lambda = 35$\,TeV, the three $\tau$-jets rate is 4.7\,fb.
For an integrated luminosity of 2\,fb$^{-1}$ (approximately what
is expected initially during Run II), this corresponds to $\sim 9$
observable events. For 30\,fb$^{-1}$, the number of observable
events is 141. The 2 $\tau$-jets cross sections of 48.1\,fb gives
$\sim 96$ events for 2\,fb$^{-1}$ of data and $\sim 1440$ for
30\,fb$^{-1}$ of data. As $\Lambda$ increases, the numbers are
smaller due to the smaller SUSY production rate. For $\Lambda =
50$\,TeV, the expected number of events for three $\tau$-jets is
about 2 for 2\,fb$^{-1}$ of data and 30 for 30\,fb$^{-1}$ of data.
The expected number of 2 $\tau$-jets events is 16 and 248 for
2\,fb$^{-1}$ and 30\,fb$^{-1}$ of data, respectively. For $\Lambda
= 70$\,TeV, the expected number of events for two $\tau$-jets is 2
and 28 for 2\,fb$^{-1}$ and 30\,fb$^{-1}$, respectively.

The branching ratios for some of the more interesting individual modes
in combined SUSY production are given in Table~\ref{spec1br}. The
electrons and muons are typically the soft to pass the cuts. Thus
requiring an $e$ or $\mu$ to enhance the signal over background
probably will be of little help.

\section{Stau NLSP line with n = 3}

\begin{figure}
\centering
\epsfxsize=0.98\textwidth
\epsfbox[76 155 501 668]{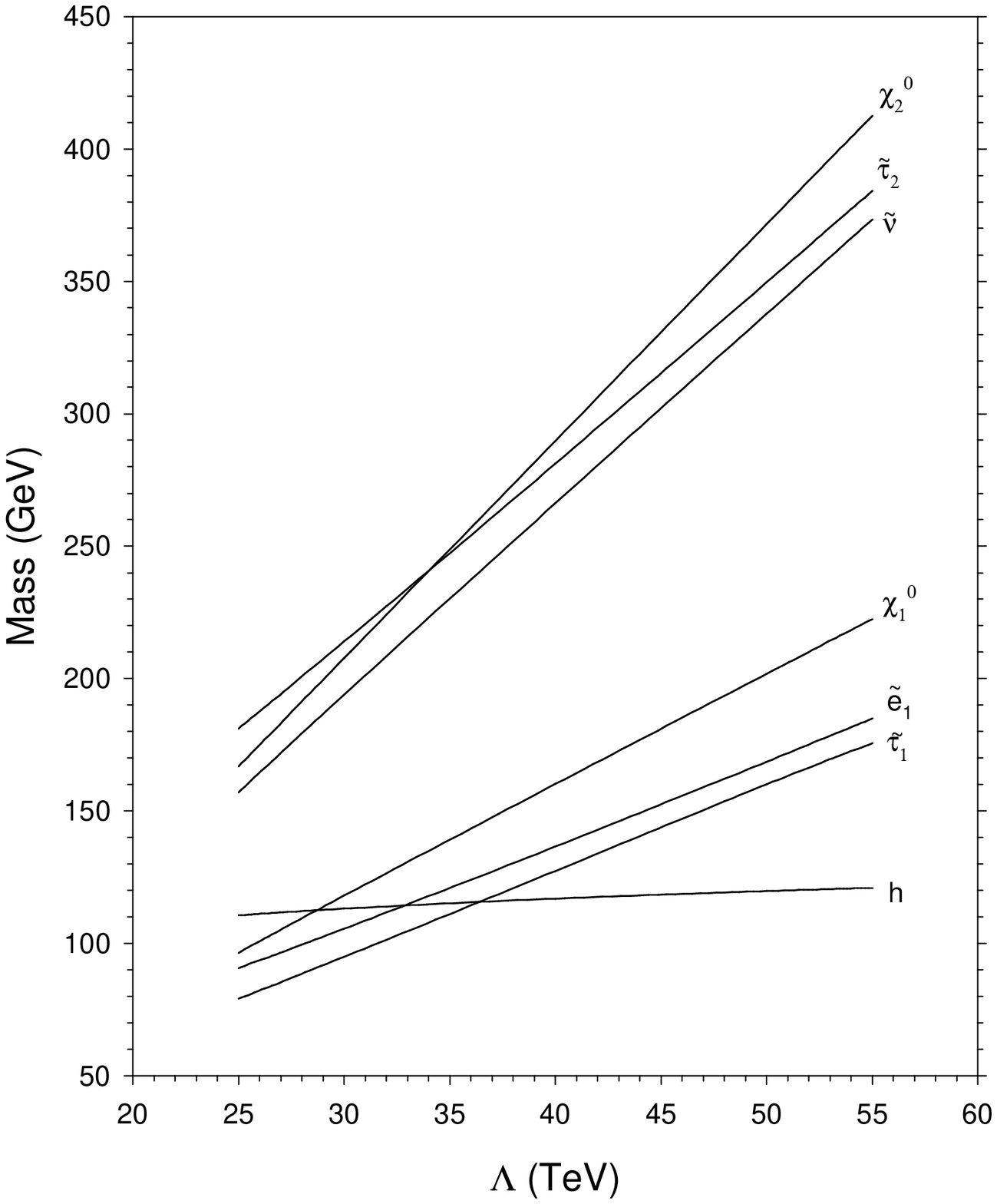}
\vskip 0.25cm
\caption{\label{mass2}
Masses for the sparticles of interest for the line
defined by $n = 3$, $\tan \beta = 15$ and $M$/$\Lambda$ = 20.
$M_{\chi_2^0} \approx M_{\chi_1^\pm}$ and
$M_{\tilde{\mu}_1} \approx M_{\tilde{e}_1}$.}
\end{figure}

We now consider a case where the ordering of the sparticles masses
is quite different from the previous case. The parameters taken
here are $n = 3$, $\tan \beta = 15$ and $M$/$\Lambda$ = 20.
We vary $\Lambda$ from 25 to 55\,TeV\@.
The masses for these parameters are given in Fig.~\ref{mass2}.
We see that the ordering of the masses is of type 2:
$M_{\chi_2^0} \agt M_{\chi_1^\pm} > m_{\tilde{\nu}} > M_{\chi_1^0}
    > m_{\tilde{e}_1, \tilde{\mu}_1} > M_{\tilde{\tau}_1}$.
This case is more complicated than the previous one due to the shifting
of the sneutrino masses below that of $\chi_1^\pm$ and $\chi_2^0$.
As a consequence, there are many decay modes for $\chi_1^\pm$ and
$\chi_2^0$ over the parameter space considered here. Moreover,
the lightest selectron and smuon masses are always below that of the
lightest neutralino. The result of all this is that the decay chains
will generally be quite involved with many steps for the values
of $\Lambda$ considered here.

The branching ratios for the sparticles are given in Table~\ref{br2}.
Since the masses of the lightest selectron and the lightest smuon are always
below that of the lightest neutralino, there are three decay modes
available for the values of $\Lambda$ considered:
$\chi_1^0 \rightarrow \tilde{\tau}_1 \tau$,
$\chi_1^0 \rightarrow \tilde{\mu}_1 \mu$ and
$\chi_1^0 \rightarrow \tilde{e}_1 e$. The decay to the stau is the
dominant decay mode especially at low values of $\Lambda$.

There are many potential decay modes for the chargino with these values of
the parameters. Since the sneutrinos are now less massive than the
chargino, these provide three decay modes that were not present in the
previous case: $\chi_1^\pm \rightarrow \tilde{\nu}_\tau \tau$,
$\chi_1^\pm \rightarrow \tilde{\nu}_\mu \mu$ and
$\chi_1^\pm \rightarrow \tilde{\nu}_e e$.
For the entire range of parameters considered, these decays to the sneutrinos
are always present as well as the decay
$\chi_1^\pm \rightarrow \tilde{\tau}_1 \nu_\tau$ which is the dominant decay
mode for $\Lambda$ less than about 50\,TeV\@. As $\Lambda$ increases, the
mass difference between the lightest chargino and the lightest neutralino
increases and the decay $\chi_1^\pm \rightarrow \chi_1^0 \, W$ becomes
kinematically allowed. At $\Lambda = 55$\,TeV, it is as important a decay
mode as $\chi_1^\pm \rightarrow \tilde{\tau}_1 \nu_\tau$. One other
distinguishing characteristic of this case from the last one is that the
heavier selectron and smuon have masses below that of $\chi_1^\pm$ and
$\chi_2^0$. Thus the decays $\chi_1^\pm \rightarrow \tilde{\tau}_2 \nu_\tau$,
$\chi_1^\pm \rightarrow \tilde{\mu}_2 \nu_\mu$ and
$\chi_1^\pm \rightarrow \tilde{e}_2 \nu_e$ are available and their
branching ratios are small, but not unimportant at large $\Lambda$.

For the second lightest neutralino, there are up to eleven main decay modes.
For low values of $\Lambda$ the dominant decay mode is
$\chi_2^0 \rightarrow \tilde{\tau}_1 \tau$. The other slepton decay modes
$\chi_2^0 \rightarrow \tilde{e}_1 e$ and
$\chi_2^0 \rightarrow \tilde{\mu}_1 \mu$ are also important. As
$\Lambda$ increases, the decay $\chi_2^0 \rightarrow \chi_1^0 h$ becomes
kinematically allowed and rapidly becomes the dominant decay. The decay
modes to the sneutrinos also become more important.

\begin{figure}
\centering
\epsfxsize=0.6\textwidth
\epsfbox[82 231 488 544]{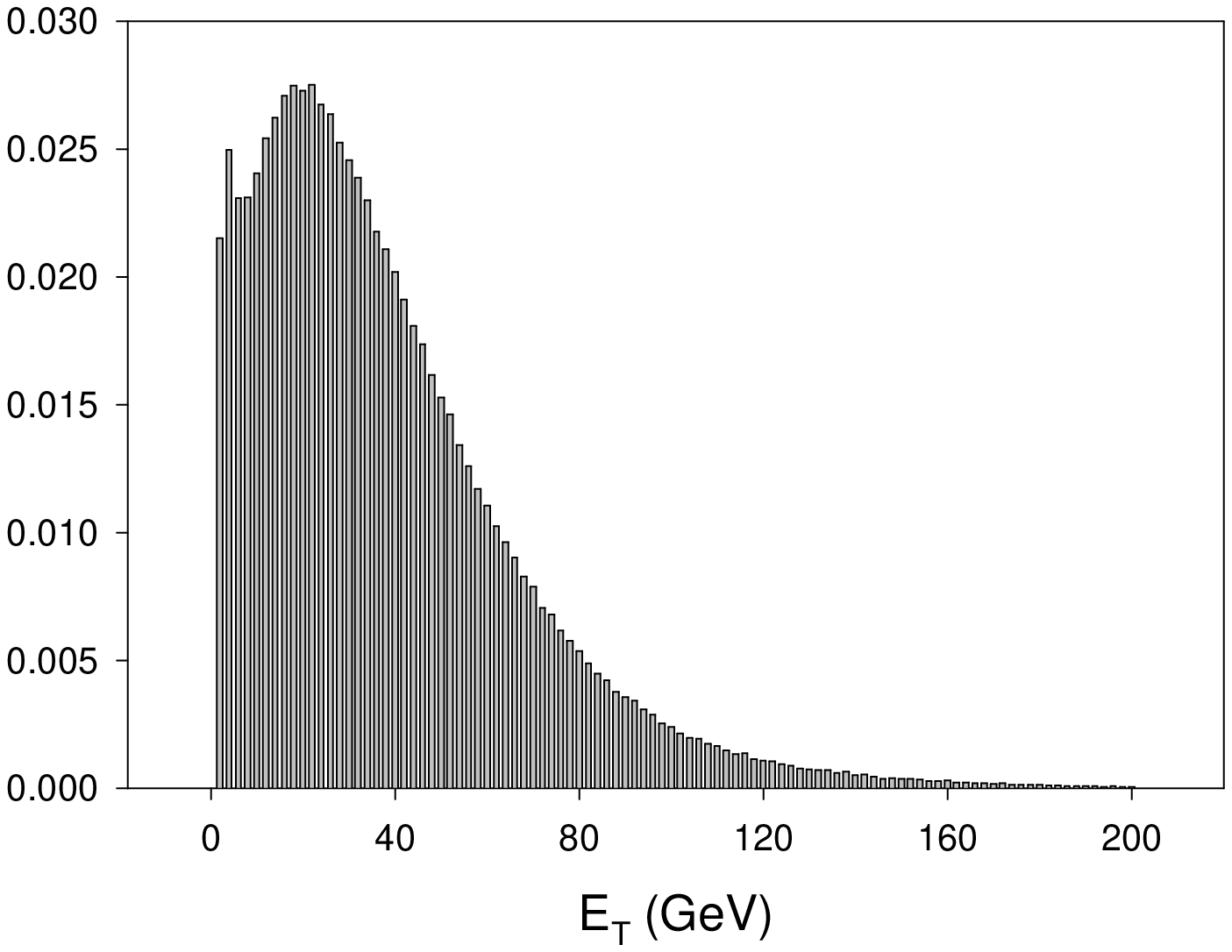}
\begin{minipage}{0.6\textwidth}
(b)
\end{minipage}
\vspace{1.5 cm}
\caption{\label{let2} The $E_T$ distributions of the leading $\tau$ jet
for the parameters $n = 3$, $\tan\beta = 15$, $M$/$\Lambda$ = 20 and
$\Lambda = 25$\,TeV\@.
In (a), no cuts have been imposed.
In (b), the $|\eta| < 1$ cut has been imposed on the $\tau$-jets.}
\end{figure}

\begin{figure}
\centering
\epsfxsize=0.6\textwidth
\epsfbox[84 232 497 544]{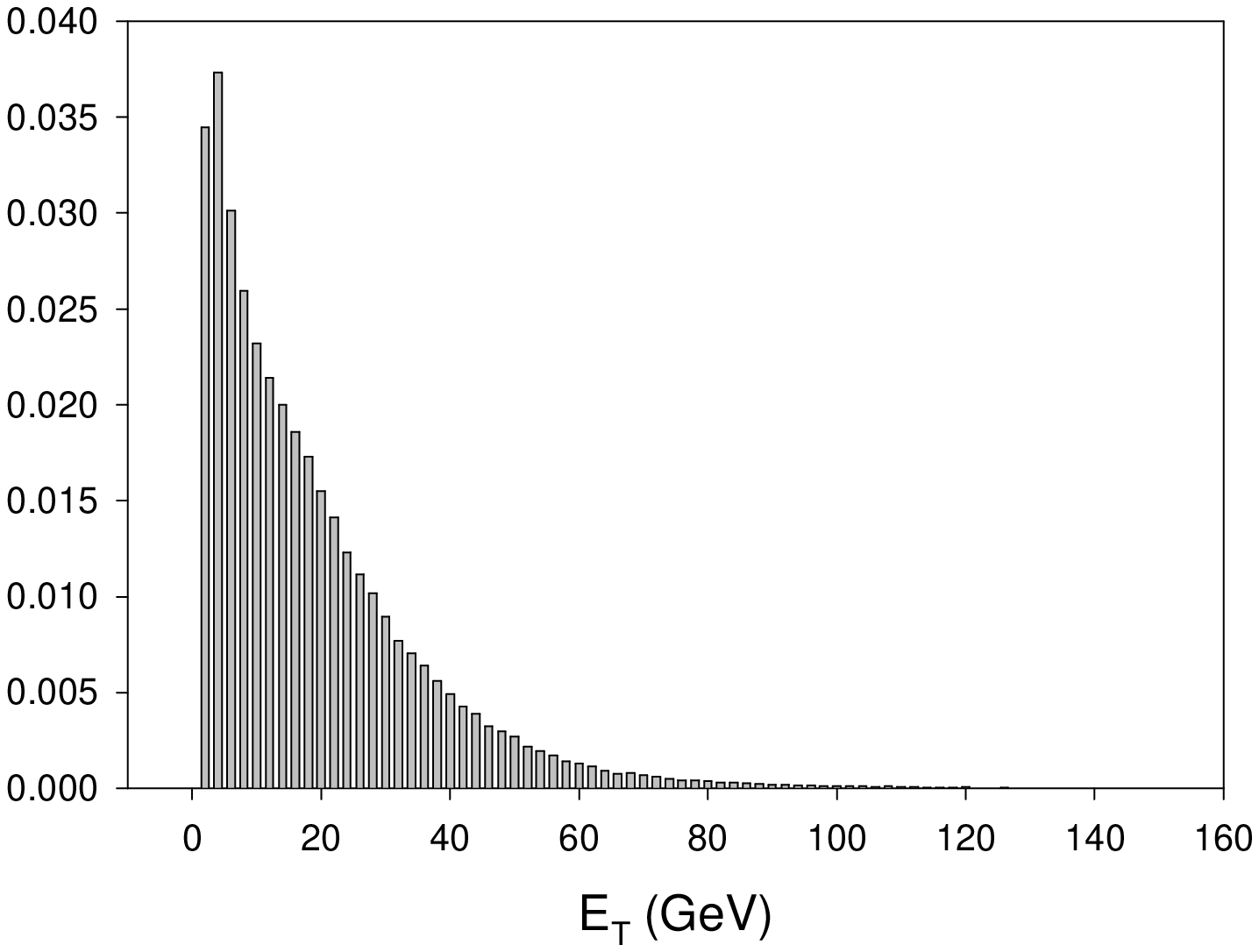}
\begin{minipage}{0.6\textwidth}
(b)
\end{minipage}
\vspace{1.5 cm}
\caption{\label{set2} The $E_T$ distributions of the secondary $\tau$ jet
for the parameters $n = 3$, $\tan\beta = 15$, $M$/$\Lambda$ = 20 and
$\Lambda = 20$\,TeV\@.
In (a), no cuts have been imposed.
In (b),  $|\eta| < 1$ cut on $\tau$-jets has been imposed.}
\end{figure}

\begin{figure}
\centering
\epsfxsize=0.6\textwidth
\epsfbox[82 232 487 545]{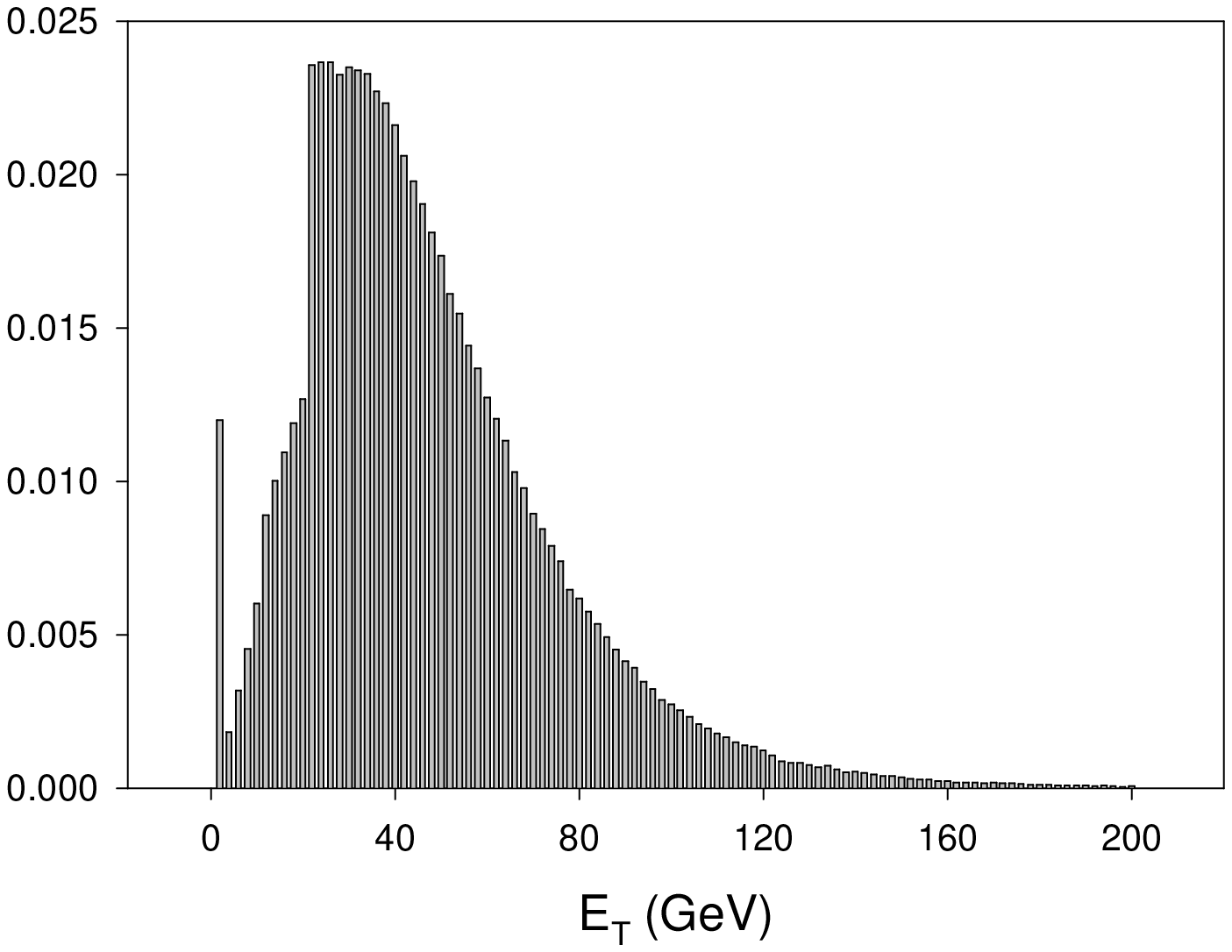}
\begin{minipage}{0.6\textwidth}
(b)
\end{minipage}
\vspace{1.5 cm}
\caption{\label{met2} \met\ distribution of the secondary $\tau$ jet for
the parameters $n = 3$, $\tan\beta = 15$, $M$/$\Lambda$ = 20 and
$\Lambda = 25$\,TeV\@.
In (a), no cuts have been imposed.
In (b), the $E_T$/$p_T$ and pseudorapidity cuts on the jets and charged
leptons have been imposed.}
\end{figure}

The $E_T$ distribution of the leading $\tau$-jet for
$\Lambda = 25$\,TeV is given in
Fig.~\ref{let2} and the $E_T$ distribution of the secondary $\tau$-jet
is given in Fig.~\ref{set2}. The distribution for the leading $\tau$-jet
is quite similar to the previous case, but the secondary $\tau$-jet
spectrum is softer due to the decrease in the direct production of
$\tau$ leptons from chargino and neutralino decays and more of the $\tau$
leptons coming from further down the decay chain. The \met\
distribution is given in Fig.~\ref{met2}.

We now consider the details of the various final state possiblities.
Table~\ref{tau2br25} gives the inclusive branching ratios for different
numbers of $\tau$-jets for $\Lambda = 25$\,TeV\@. In principle, up to
six $\tau$ leptons can be produced in $\chi_1^+ \chi_1^-$ production,
but the five and six $\tau$ lepton branching ratios are small. The most
important mode before cuts is the two $\tau$-jet mode at 40.4\%, but the
one and three $\tau$-jets modes are also appreciable. After implementing
the cuts, the branching ratios are greatly decreased: the two $\tau$-jets
branching ratio becomes only 11.4\% and the one $\tau$-jet branching ratio
becomes 23.4\%. The three $\tau$-jet branching ratio becomes essentially
negligible. The situation changes as $\Lambda$ increases.
Table~\ref{tau2br40} gives the inclusive $\tau$-jet branching ratios for
$\Lambda = 40$\,TeV\@. The branching ratio for greater numbers of
$\tau$-jets are now larger. This is due to the decrease in the branching
ratio for $\chi_1^\pm \rightarrow \tilde{\tau}_1 \nu_\tau$ from which one
can get only one $\tau$-jet from the chargino and the increase in
$\chi_1^\pm \rightarrow \chi_1^0 W$ which can give three $\tau$-jets and
$\chi_1^\pm \rightarrow \tilde{\nu}_\tau \tau$ which can also give
three $\tau$-jets. The two $\tau$-jets mode in $\chi_1^+ \chi_1^-$
production is still dominant at 34.8\%, but now the three $\tau$-jets mode
is appreciable at 28.6\%. After cuts, the three $\tau$-jets rate drops
to 5\% and the one $\tau$-jet mode becomes dominant at 30.1\%.

Turning now to $\chi_2^0 \, \chi_1^\pm$ production, we see that at
low $\Lambda$, there is the potential to produce up to five
$\tau$-jets (three from $\chi_1^\pm \rightarrow \tilde{\nu}_\tau
\tau$ with the subsequent decays $\tilde{\nu}_\tau \rightarrow
\chi_1^0 \, \nu_\tau$ and $\chi_1^0 \rightarrow \tilde{\tau}_1
\tau$ and two $\tau$-jets from $\chi_2^0 \rightarrow
\tilde{\nu}_\tau \nu_\tau$), but the branching ratios for more
than three $\tau$-jets are rather small. As usual the dominant
decay mode is to two $\tau$-jets with a before cuts branching
ratio of 40.8\%, but the three $\tau$-jets branching ratio is
large at 29.1\% as shown in Table~\ref{tau2br25}.
 After cuts these fall to 16.8\%
and 3.3\%, respectively. The one $\tau$-jet mode becomes dominant
with a branching ratio of 24.5\%. Table~\ref{tau2br40} gives the
results for $\Lambda = 40$\,TeV\@. The four $\tau$-jets mode has a
substantial decay rate before cuts, but this mode becomes
negligible after cuts due to the softness of the fourth $\tau$-jet
due to its production further down the decay chain. On the other
hand, the three $\tau$-jets branching ratio is now higher at
6.7\%.

Slepton production for this case is largely the same as in the previous
case. With $\tilde{\tau}_1 \rightarrow \tau \, \tilde{G}$ being
essentially the only decay mode for the lighter stau, the $\tau$-jet
branching ratios before cuts are completely dictated by the hadronic
branching ratio for the $\tau$ lepton.
For $\tilde{\mu}_1^+ \tilde{\mu}_1^-$ and $\tilde{e}_1^+ \tilde{e}_1^-$
production, up to four $\tau$-jets can be produced, but after cuts
the rates for three and four $\tau$-jets are greatly reduced.

\begin{figure}
\centering
\epsfxsize=0.98\textwidth
\epsfbox[73 172 499 679]{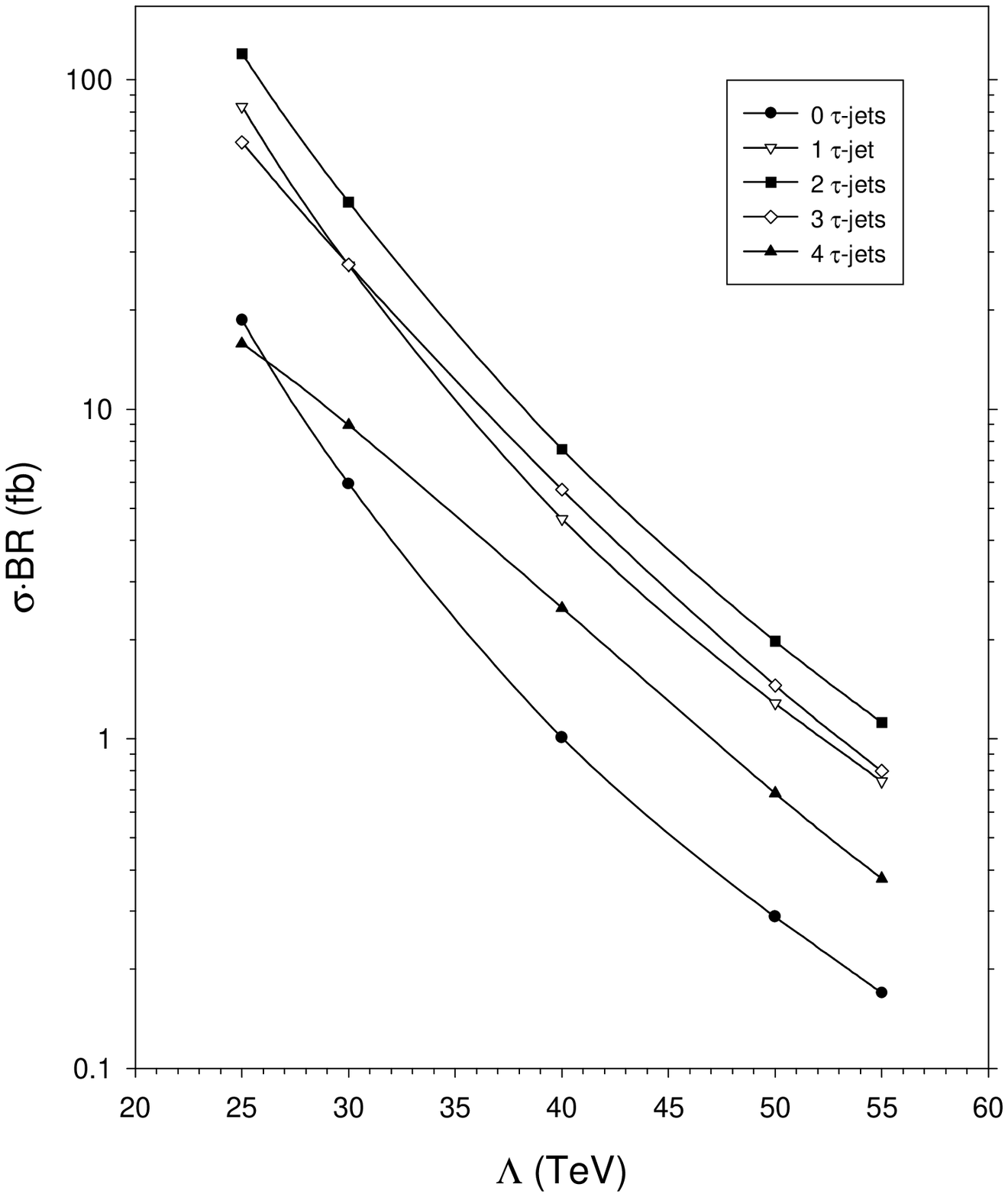}
\vskip 0.25cm
\caption{\label{tauj2nc} $\sigma \cdot BR$ before cuts for the inclusive
$\tau$ jets modes
for the parameters $n = 3$, $\tan \beta = 15$ and $M$/$\Lambda$ = 20.}
\end{figure}

\begin{figure}
\centering
\epsfxsize=0.98\textwidth
\epsfbox[70 159 500 662]{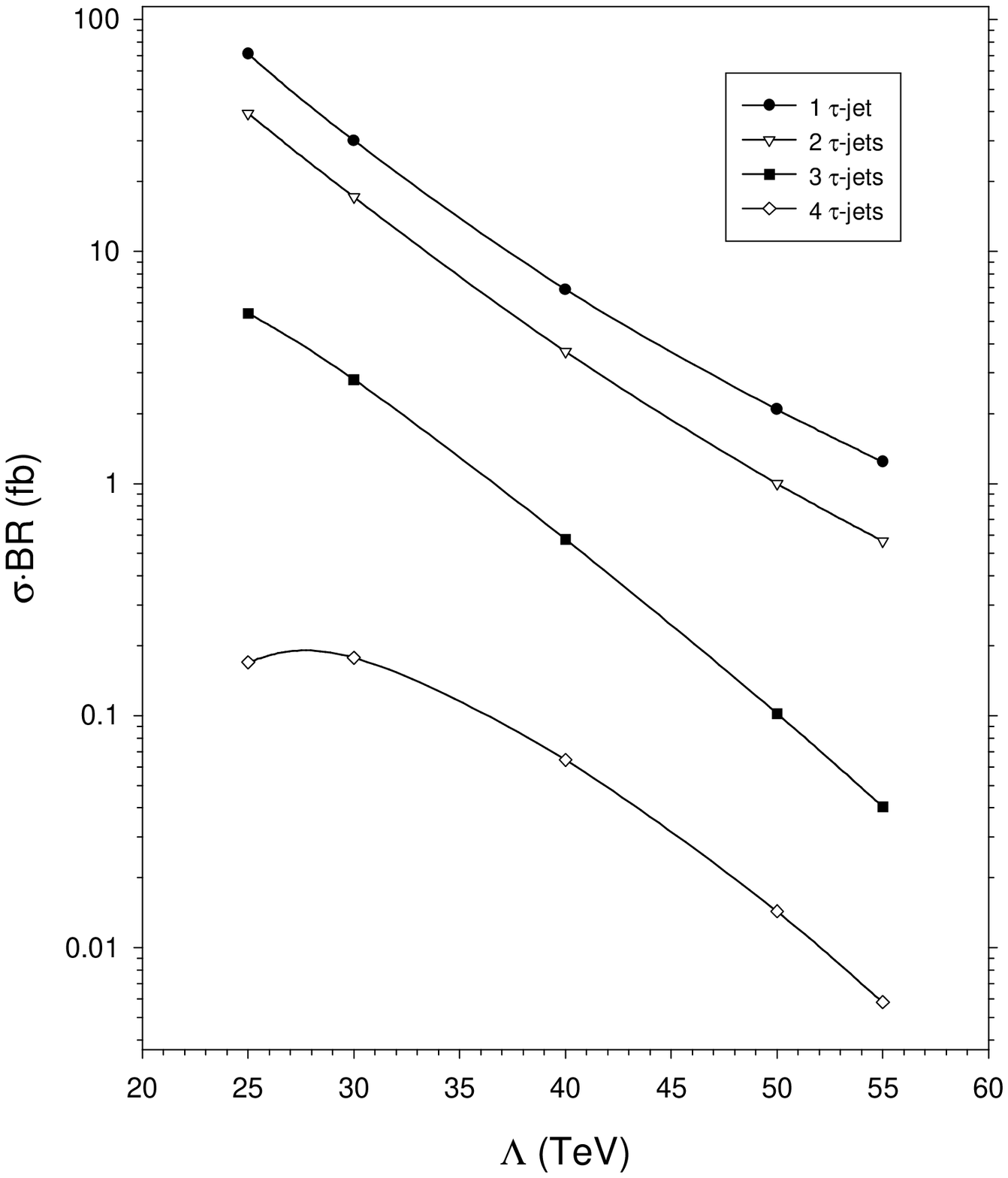}
\vskip 0.25cm
\caption{\label{tauj2c} $\sigma \cdot BR$ after cuts for the inclusive
$\tau$ jets modes
for the parameters $n = 3$, $\tan \beta = 15$ and $M$/$\Lambda$ = 20.}
\end{figure}

Putting all the pieces together, we can now answer the question as
to the probability of observing these events at Tevatron's Run II
and Run III\@. Fig.~\ref{tauj2nc} shows the branching ratios for
the inclusive $\tau$-jet modes before cuts for all the considered
SUSY production modes combined. The two $\tau$-jets mode is the
mode with the largest $\sigma \cdot {\rm BR}$, but the production
rates for one and three $\tau$-jets are close to this. After
including cuts, the one $\tau$-jet mode is dominant and the two
$\tau$-jets mode is respectably high as seen in Fig.~\ref{tauj2c}.
For $\Lambda = 25$\,TeV, the three $\tau$-jet rate is 5.4\,fb
giving $\sim 11$ for 2\,fb$^{-1}$ of data and $\sim 162$ events
for 30\,fb$^{-1}$ of data. The two $\tau$-jet rate of 39.3\,fb
gives $\sim 79$ and $\sim 1179$ events, respectively. For the
higher $\Lambda$ value of 50\,TeV, the rates are cut down
significantly. The two $\tau$-jets rate of 1.0\,fb gives $\sim 2$
and $\sim 30$ events for 2\,fb$^{-1}$ and 30\,fb$^{-1}$ of data,
respectively.

\section{Co-NLSP Case}

\begin{figure}
\centering
\epsfxsize=0.98\textwidth
\epsfbox[73 112 500 627]{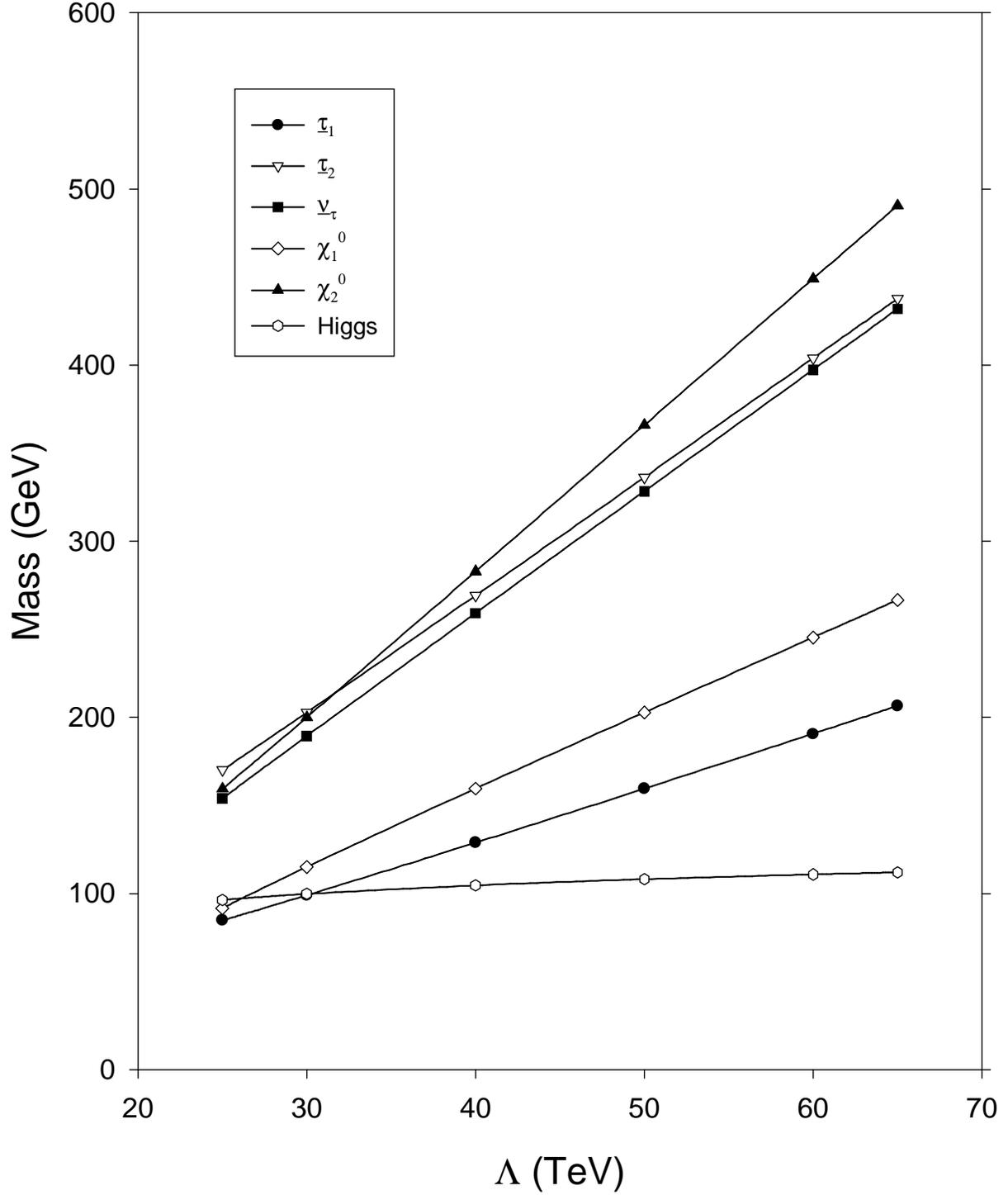}
\vskip 0.25cm
\caption{\label{comass} The masses for the sparticles of interest for
the co-NLSP example where $n = 3$, $\tan \beta = 3$ and
$M$/$\Lambda$ = 3.}
\end{figure}

\begin{figure}
\centering
\epsfxsize=0.98\textwidth
\epsfbox[60 133 493 545]{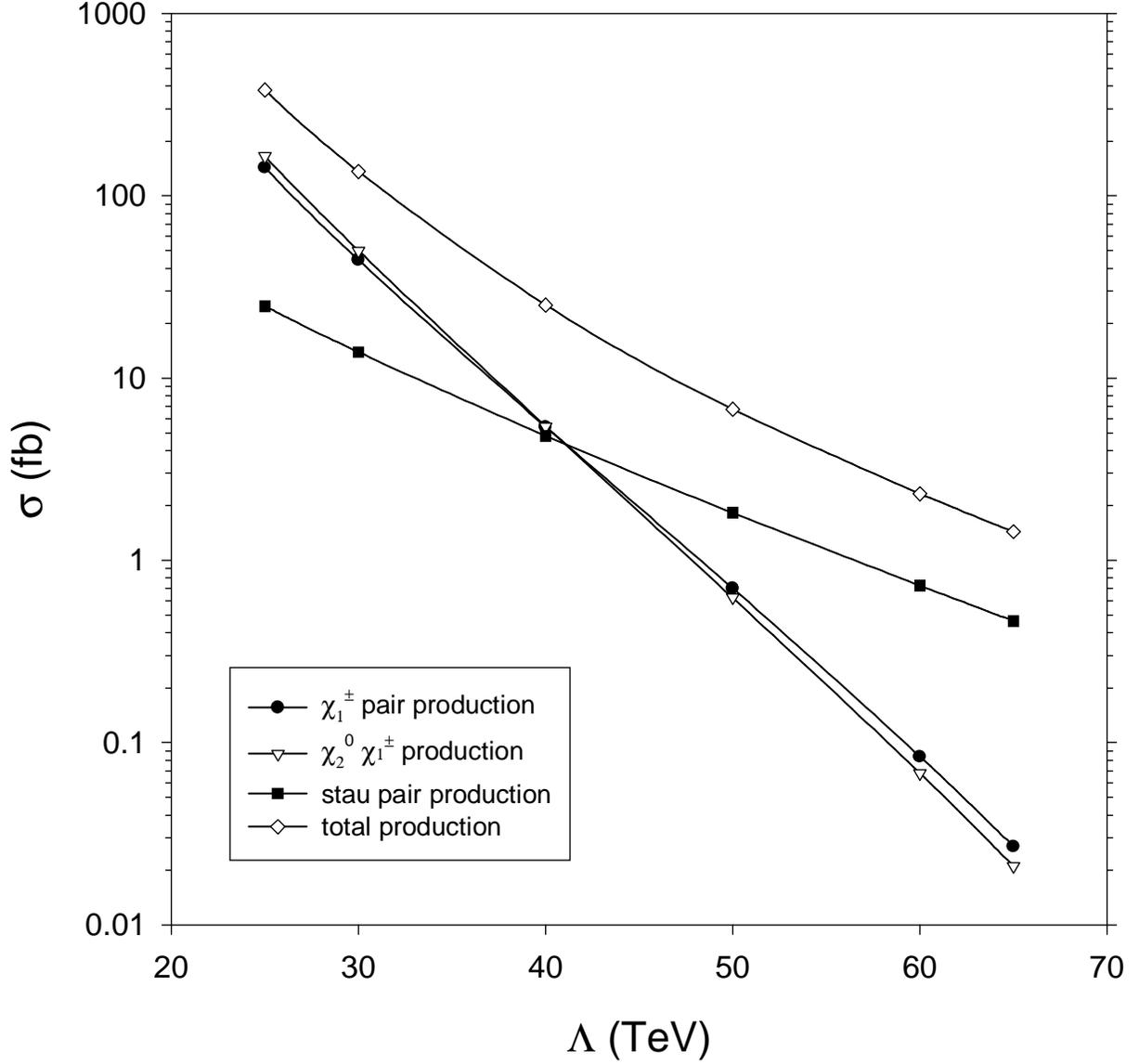}
\vskip 0.25cm
\caption{\label{co_cross}
The SUSY production cross sections for
the co-NLSP example where $n = 3$, $\tan \beta = 3$ and
$M$/$\Lambda$ = 3.}
\end{figure}

The co-NLSP case \cite{akm} refers to when the mass difference between
$\tilde{e}_1$ (and $\tilde{\mu}_1$) and $\tilde{\tau}_1$ is less than the
mass of the $\tau$ lepton. When this is the case, the three-body
decay mode $\tilde{e}_1 \rightarrow e \tau \tilde{\tau}_1$
is not kinematically
allowed and the main decay mode for the selectron is the two-body mode
$\tilde{e}_1 \rightarrow e \, \tilde{G}$. The parameters for the example of
this case that is considered here are $n = 3$, $\tan \beta = 3$ and
$M$/$\Lambda$ = 3. $\Lambda$ is varied from 25 to 65\,TeV\@.
The masses of the sparticles of interest are given in Fig.~\ref{comass}.
We see that the oredering of the masses here is a special case of type 2:
$M_{\chi_2^0} \agt M_{\chi_1^\pm} > m_{\tilde{\nu}} > M_{\chi_1^0} >
m_{\tilde{e}_1, \tilde{\mu}_1} \approx m_{\tilde{\tau}_1}$. With this
ordering of the masses there are typically many decays mode of the
sparticles to consider. The cross sections for the SUSY production modes
are given in Fig.~\ref{co_cross}. Due to the rapid increase in the
sparticle masses (especially the gaugino masses) as $\Lambda$  is
increased, the cross sections tend to decrease fairly rapidly.

The branching ratios for the sparticles of interest are given in
Table~\ref{co_br}. Since the lighter selectron and the lighter smuon
have about the same mass as the stau, the branching ratios for
$\chi_1^0 \rightarrow \tilde{\tau}_1 \tau$,
$\chi_1^0 \rightarrow \tilde{\mu}_1 \mu$ and
$\chi_1^0 \rightarrow \tilde{e}_1 e$ are nearly equal.
The decay to the stau is slightly favored.

The chargino's decays strongly depend on the value of $\Lambda$. For
values of $\Lambda$ that are 30\,TeV and below, the dominant decay mode
of the chargino is $\chi_1^\pm \rightarrow \tilde{\tau}_1 \nu_\tau$. As
$\Lambda$ is increased above this, the chief decay modes are the decays
to the sneutrinos and the decay $\chi_1^\pm \rightarrow \chi_1^0 W$
which tends to dominate when kinematically allowed. As $\Lambda$
increases above 30\,TeV, the masses of the heavier sleptons
($\tilde{\tau}_2$, $\tilde{\mu}_2$ and $\tilde{e}_2$) fall below the
mass of the chargino  and so the decays to these heavier sleptons are
allowed as well.

For the second lightest neutralino, there are again up to 11 main decay
modes. At low values of $\Lambda$, the decays to the lighter sleptons
are dominant with $\chi_2^0 \rightarrow \tilde{\tau}_1 \tau$ having a
slight edge over the other two slepton decays. As $\Lambda$ increases
the decays to the sneutrinos gradually become more important. In
addition the decay $\chi_2^0 \rightarrow \chi_1^0 \, {\rm h}$ and the decays
to the heavier sleptons become kinematically allowed
and dominate over the other decays.

In chargino pair production at low values of $\Lambda$,
two $\tau$ leptons are always produced because essentially the only
decay mode for the chargino is
$\chi_1^\pm \rightarrow \tilde{\tau}_1 \nu_\tau$ while the stau
decays via $\tilde{\tau}_1 \rightarrow \tau \, \tilde{G}$.
On the other hand, the classic three $\tau$ signature for
$\chi_2^0 \, \chi_1^\pm$ production will be diminished since
$\chi_2^0 \rightarrow \tilde{\tau}_1 \tau$,
$\chi_2^0 \rightarrow \tilde{\mu}_1 \mu$ and
$\chi_2^0 \rightarrow \tilde{e}_1 e$ are all roughly equal. Since the
subsequent decays of the selectron and smuon to the gravitino produces
no $\tau$ leptons (unlike the three-body decay modes
$\tilde{e}_1 \rightarrow e \tau \tilde{\tau}_1$ and
$\tilde{\mu}_1 \rightarrow \mu \tau \tilde{\tau}_1$ that were
dominant in the other two cases), there will tend to be a depletion in
$\tau$-jets here relative to the previous type 2 case which
didn't satisfy the co-NLSP condition. For larger values of $\Lambda$, the
situation is more complicated, but the decay will frequently involve
the lightest neutralino. The lightest neutralino in turn tends to decay
to $\tilde{\tau}_1$, $\tilde{\mu}_1$ and $\tilde{e}_1$ roughly equally.
Thus there is again a relative depletion in events with $\tau$-jets.

\begin{figure}
\centering
\epsfxsize=0.6\textwidth
\epsfbox[84 233 487 544]{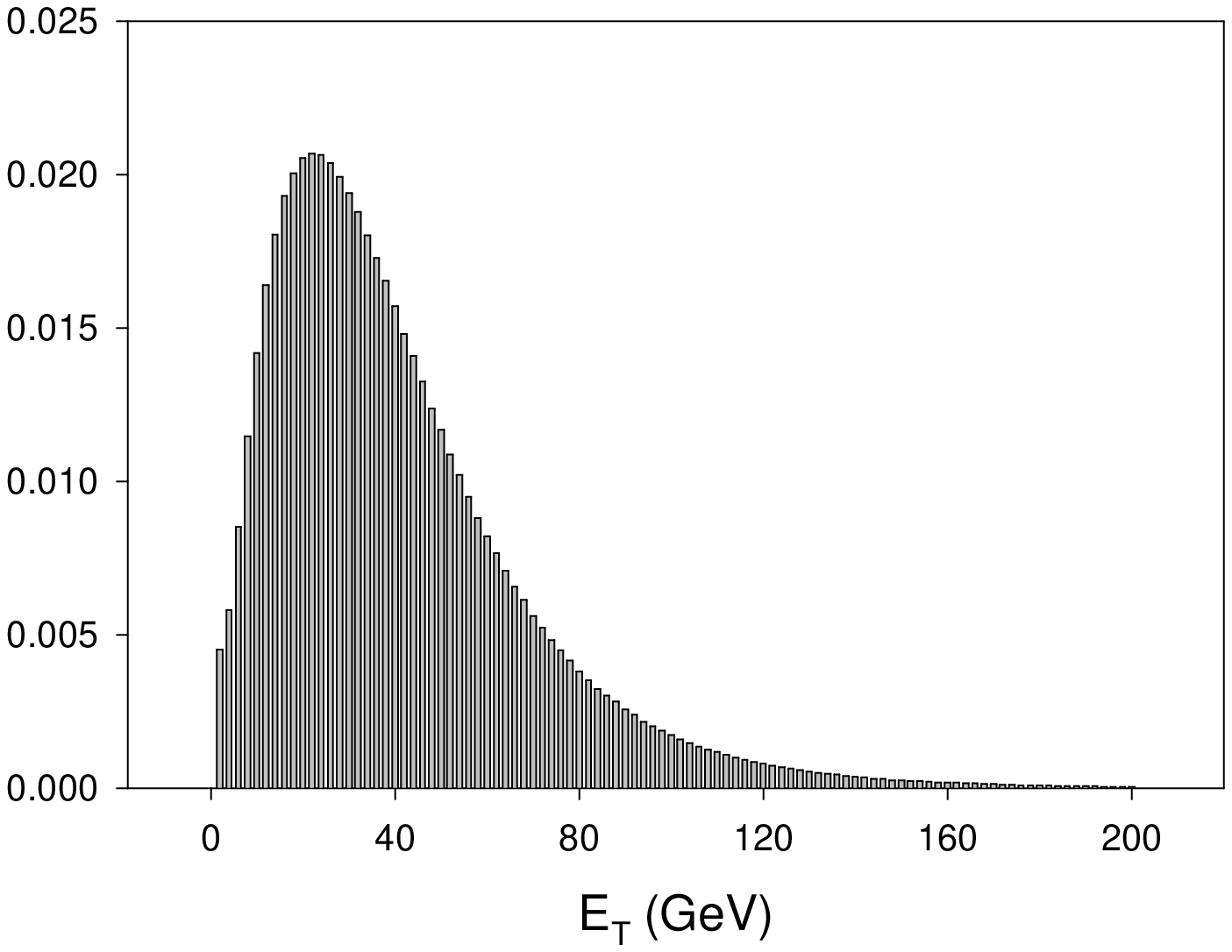}
\begin{minipage}{0.6\textwidth}
(b)
\end{minipage}
\vspace{1.5 cm}
\caption{\label{co_let} The $E_T$ distributions of the leading $\tau$ jet
for the parameters $n = 3$, $\tan\beta = 3$, $M$/$\Lambda$ = 3 and
$\Lambda = 25$\,TeV\@.
In (a), no cuts have been imposed.
In (b), the $|\eta| < 1$ cut on $\tau$-jets has been imposed.}
\end{figure}

\begin{figure}
\centering
\epsfxsize=0.6\textwidth
\epsfbox[84 232 490 544]{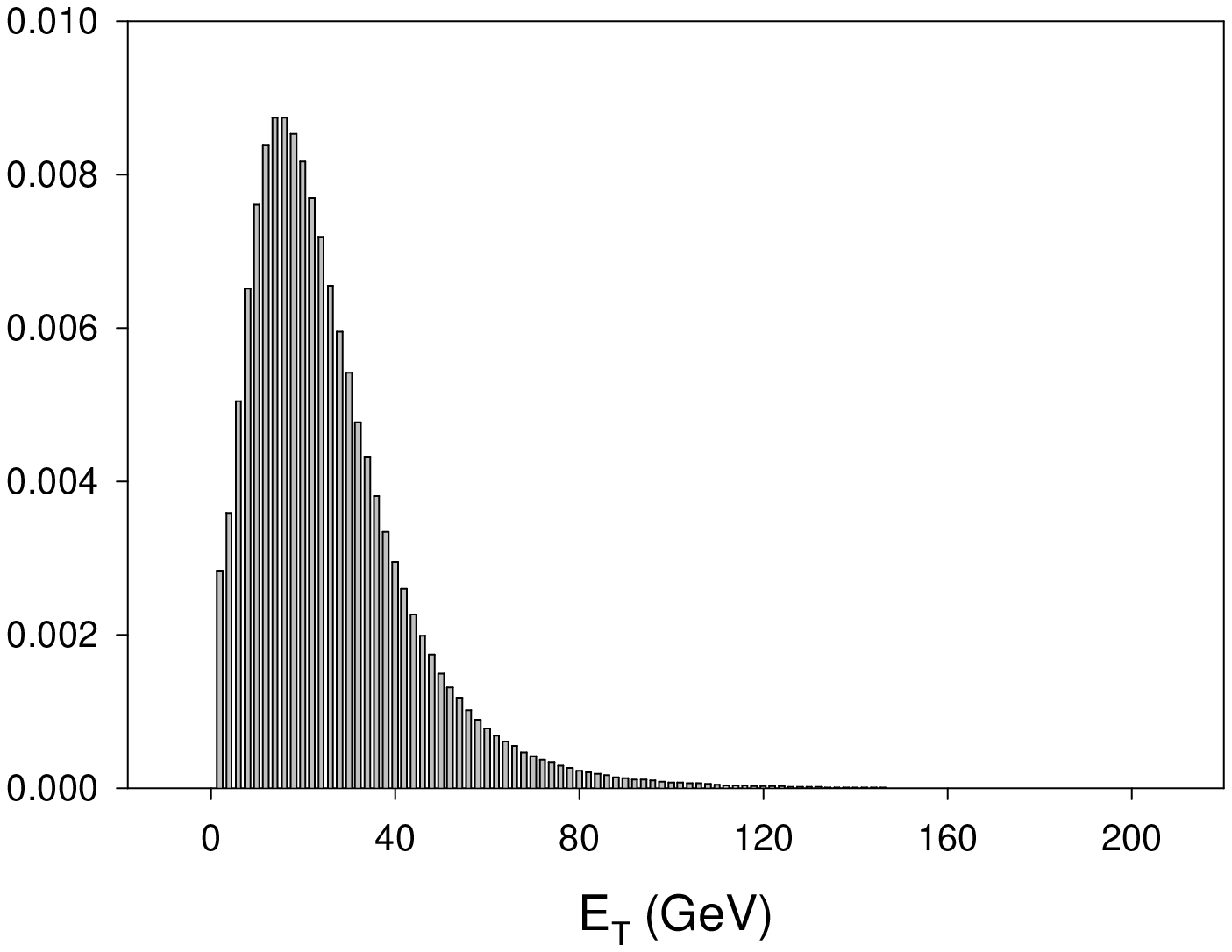}
\begin{minipage}{0.6\textwidth}
(b)
\end{minipage}
\vspace{1.5 cm}
\caption{\label{co_set} The $E_T$ distributions of the secondary $\tau$ jet
for the parameters $n = 3$, $\tan\beta = 3$, $M$/$\Lambda$ = 3 and
$\Lambda = 25$\,TeV\@.
In (a), no cuts have been imposed.
In (b), the $|\eta| < 1$ cut on $\tau$-jets has been imposed.}
\end{figure}

\begin{figure}
\centering
\epsfxsize=0.6\textwidth
\epsfbox[82 232 487 545]{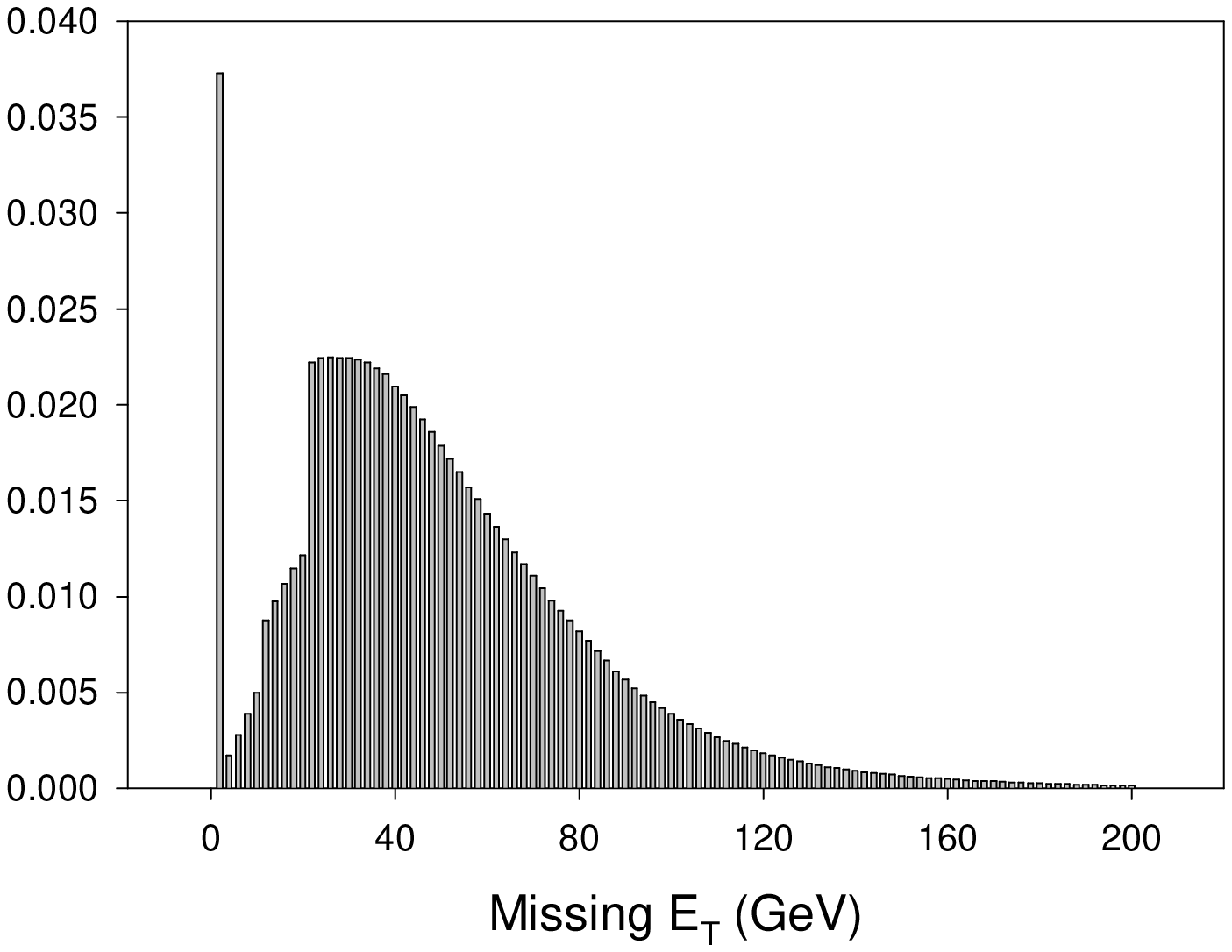}
\begin{minipage}{0.6\textwidth}
(b)
\end{minipage}
\vspace{1.5 cm}
\caption{\label{co_met} \met\ distribution of the secondary $\tau$ jet for
the parameters $n = 3$, $\tan\beta = 3$, $M$/$\Lambda$ = 3 and
$\Lambda = 25$\,TeV\@.
In (a), no cuts have been imposed.
In (b), the $E_T$/$p_T$ and pseudorapidity cuts on the jets and
charged leptons have been imposed.}
\end{figure}

The $E_T$ distribution for the leading $\tau$-jet when $\Lambda = 25$\,TeV
is given in Fig.~\ref{co_let}. The $E_T$ distribution of the secondary
$\tau$-jet is given in Fig.~\ref{co_set}. Qualitatively, these are much
the same as in the previous cases. At $\Lambda = 25$\,TeV, the decay chains
are relatively short and the $\tau$-jets tend to be quite hard. The \met\
distribution is given in Fig.~\ref{co_met}.

We now consider the details of the various final state possibilities.
Table~\ref{tau3br25} gives the inclusive branching ratios for different
numbers of $\tau$-jets for $\Lambda = 25$\,TeV\@. With
$\chi_1^\pm \rightarrow \tilde{\tau}_1 \nu_\tau$ being the only decay
mode here, $\chi_1^+ \chi_1^-$ production produces two $\tau$ leptons.
Thus the probability for $\tau$-jets before cuts  is dictated by the
hadronic branching ratio of the $\tau$ lepton. Including cuts
diminishes the number of events with a given number of $\tau$-jets.
For example, the branching ratio for 2 $\tau$-jets falls from 42\% to
10.5\%. When $\Lambda$ is increased, the situation changes dramatically.
Table~\ref{tau3br40} gives the inclusive branching ratios for
$\Lambda = 40$\,TeV\@. The possibility exists to create many $\tau$-jets,
but the probability for creating more than three is low. In addition, the
probability for producing no $\tau$-jets is high at $\sim 35$\%.
After cuts, the only appreciable modes are the two $\tau$-jets mode at 10\%
and the one $\tau$-jet mode at 18\%.

Turning now to $\chi_2^0 \, \chi_1^\pm$ production, we see that at
low $\Lambda$, there is the potential to produce up to three
$\tau$-jets. The rates are diminished by the strong presence of
$\chi_2^0 \rightarrow \tilde{e}_1 e$ and $\chi_2^0 \rightarrow
\tilde{\mu}_1 \mu$, however, and the rate for no $\tau$-jets is
high at $\sim 25$\%. After cuts, the two $\tau$-jets branching
ratio is only about 7\% and the one $\tau$-jet rate is about 25\%.
For $\Lambda = 40$\,TeV, the potential exists to create many more
$\tau$-jets, but the one and two $\tau$-jets modes remain dominant
with after cuts branching ratios of 19\% and 11\%, respectively.

For $\tilde{\tau}_1^+ \tilde{\tau}_1^-$, the situation is pretty much the
same as it is in the previous cases considered. With
$\tilde{\tau}_1 \rightarrow \tau \, \tilde{G}$ being the only decay mode
of the lighter stau, the probability for a given number of $\tau$-jets
is completely dictated by the hadronic branching ratio of the $\tau$
lepton. The branching ratios after cuts are largely dictated by the mass
of the stau.

\begin{figure}
\centering
\epsfxsize=0.98\textwidth
\epsfbox[80 132 508 644]{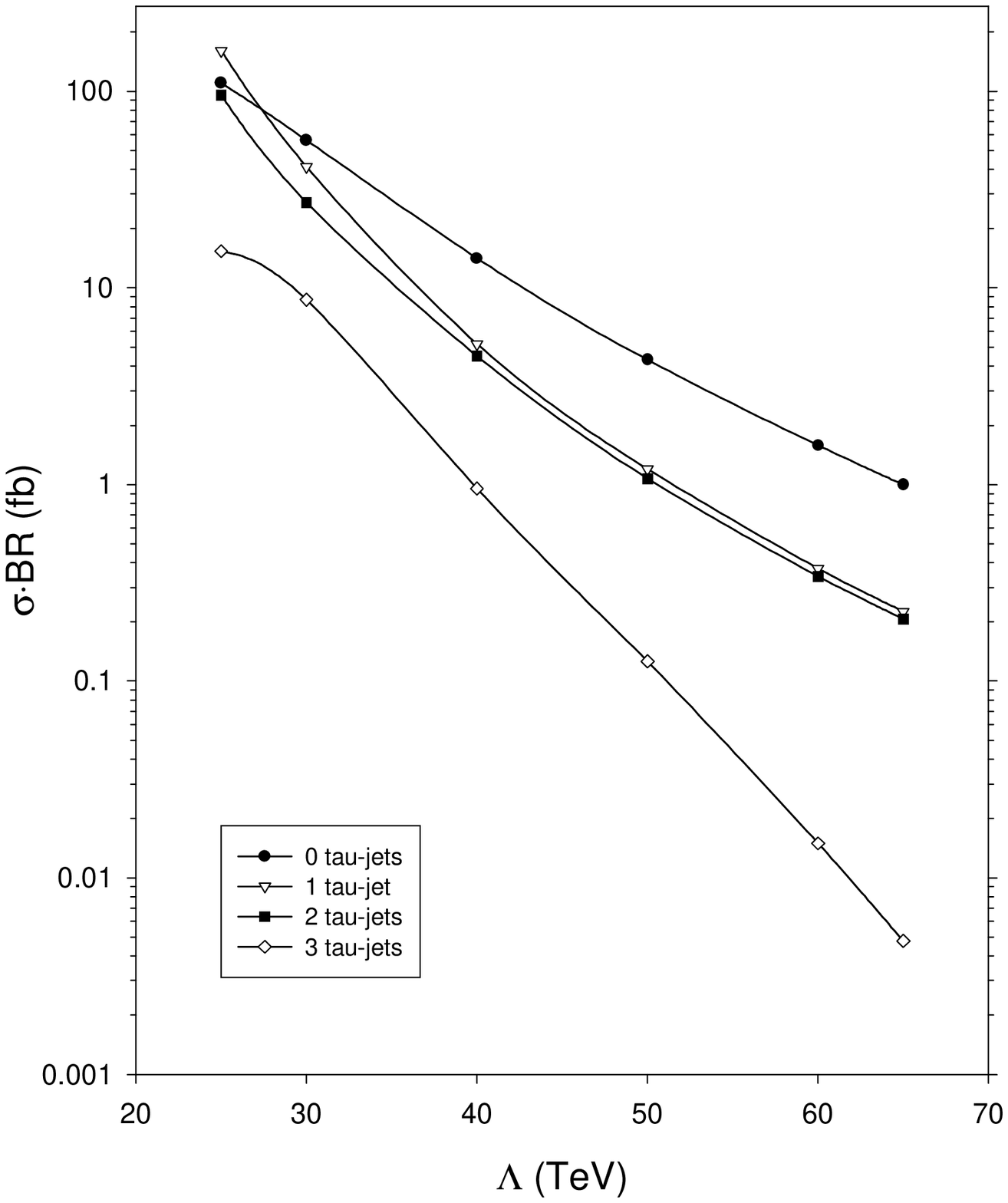}
\vskip 0.25cm
\caption{\label{tauj3nc} $\sigma \cdot BR$ before cuts for the inclusive
$\tau$ jets modes
for the parameters $n = 3$, $\tan \beta = 3$ and $M$/$\Lambda$ = 3.}
\end{figure}

\begin{figure}
\centering
\epsfxsize=0.98\textwidth
\epsfbox[95 135 524 588]{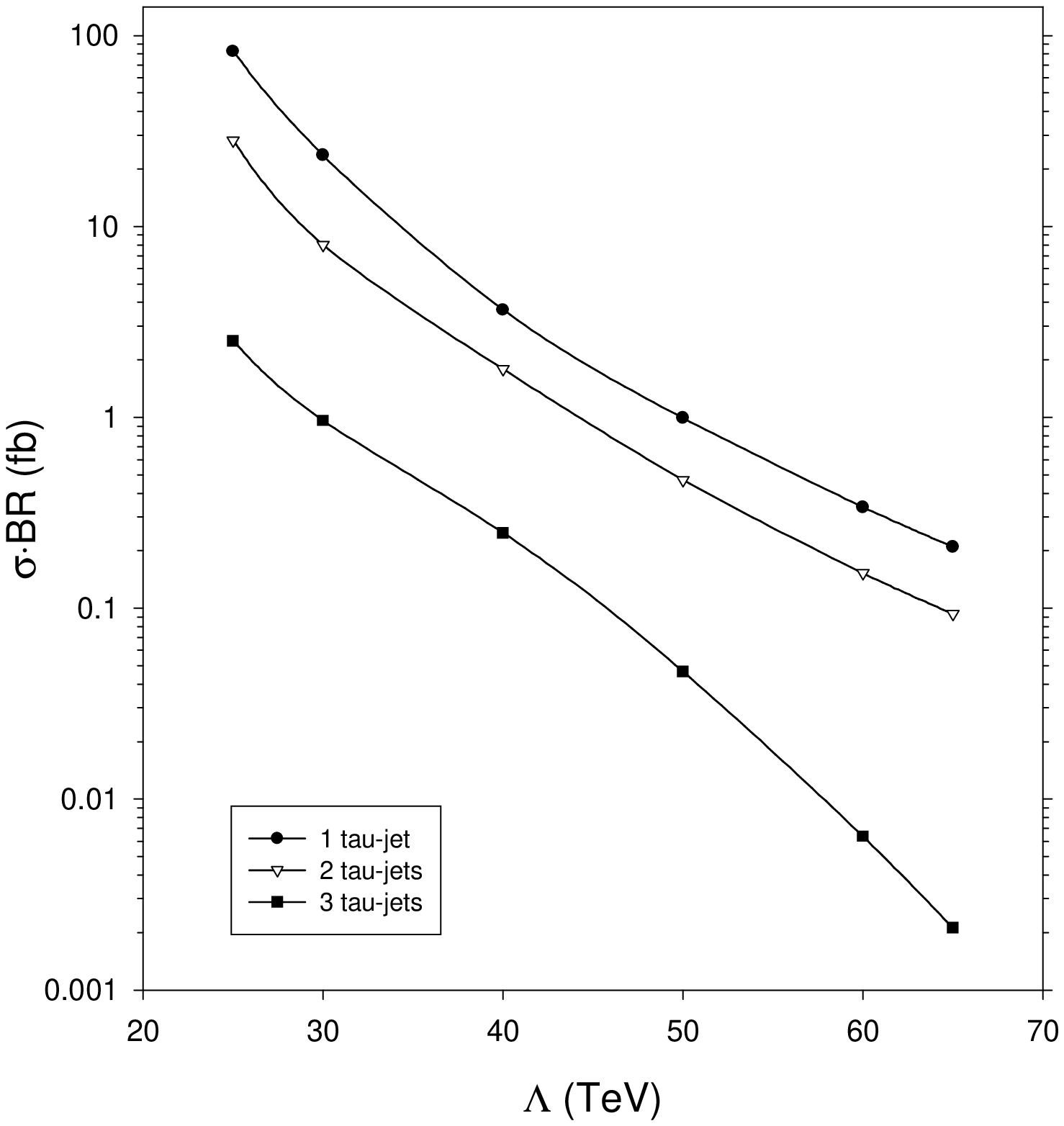}
\vskip 0.25cm
\caption{\label{tauj3c} $\sigma \cdot BR$ after cuts for the inclusive
$\tau$ jets modes
for the parameters $n = 3$, $\tan \beta = 3$ and $M$/$\Lambda$ = 3.}
\end{figure}

We now consider the possibility of observing these events at the Tevatron's
Run II and TeV33. Fig.~\ref{tauj3nc} shows the combined production rates for
the inclusive $\tau$-jet modes before cuts for all the SUSY production
modes considered. We do not include the cross sections for more than three
$\tau$-jets as these are prohibitively small. In sharp contrast to the
previous cases, the most typical situation is that no $\tau$-jets are
produced. For low values of $\Lambda$ ($\Lambda < 40$\,TeV), however,
the production rates for one and two $\tau$-jets are comparable.
The results after cuts are shown in Fig.~\ref{tauj3c}. The one $\tau$-jet
mode is dominant and the two $\tau$-jet mode is respectably high. For
$\Lambda = 25$\,TeV, the two $\tau$-jets rate is about 28 fb which gives
about 56 events for 2\,fb$^{-1}$ of data and 840 events for 30\,fb$^{-1}$
of data. For $\Lambda = 40$\,TeV, we have a lower production rate of
1.8\,fb. For 2\,fb$^{-1}$ of data this corresponds to $\sim 4$ events,
while 30\,fb$^{-1}$ of data gives about 54 events.

The branching ratios for some of the more important individual modes
are given in Table~\ref{spec3br}. Unlike the previous cases considered,
there is the potentialthat modes with specific numbers of charged
leptons could be important. We see from the table that for
$\Lambda = 25$\,TeV, the rate for an electron and a $\tau$-jet is 12.7\,fb
after cuts. For electrons and muons combined, this is 25.4\,fb. For an
integrated luminosity of 2\,fb$^{-1}$, this corresponds to about 50 events,
while for an integrated luminosity of 30\,fb$^{-1}$, this corresponds to
about 762 events. A better signal is the 2 $e$ + $\tau$-jet and
2 $\mu$ + $\tau$-jet signals. The combined cross section for this is 13\,fb.
For 2\,fb$^{-1}$ of data, this corresponds to 26 events.
For 30\,fb$^{-1}$ of data, this corresponds to about 390 events.

\section{Conclusion}

We have considered the phenomenology of GMSB models where the lighter
stau is the NLSP and decays promptly. We have looked at a wide range of
the GMSB parameter space. Typically, in those regions of the parameter
space where SUSY production can occur at an observable rate, the dominant
SUSY production modes are $\chi_1^+ \, \chi_1^-$ and
$\chi_2^0 \, \chi_1^\pm$, although slepton pair production can be
significant in those regions of the parameter space where $\chi_1^\pm$ and
$\chi_2^0$ are too massive to be readily produced. The decays of the
SUSY particles lead to events containing two or three $\tau$ leptons
plus large missing transverse energy. Searching for the $\tau$ lepton
signals by the hadronic decays of the $\tau$ leptons to thin jets is
complicated by the fact that, while primary $\tau$-jets can have high
$E_T$, the secondary $\tau$-jets tend to be rather soft. As a result,
many of the $\tau$-jets tend to be eliminated by the cuts. We've shown
that the most promising channel is the two $\tau$-jets mode, while
the three $\tau$-jets mode can be important at the higher integrated
luminosity at Run III\@. This $\tau$-jet is degraded in the co-NLSP
case where the lighter charged sleptons are nearly degenerate in mass.
The missing transverse energy associated with the events is large
providing a good trigger for these events. Good $\tau$ identification
will be extremely important to detect the signal.

\section*{Acknowledgments}

We thank B.~Dutta and G.~Wolf for many helpful discussions and S.~P.~Martin
for some useful correspondence. This research was supported by U.S.
Department of Energy Grant No. DE-FG03-98ER41076.

\newpage

\begin{table}
\centering
\caption{\label{br1} Branching ratios of the sparticles of interest for the
parameters $n = 2$, $\tan \beta = 15$ and $M$/$\Lambda$ = 3. The decays
of the $\tilde{\mu}_1$ are obtained by replacing the $e$ with a $\mu$ in
the $\tilde{e}_1$ decays.}
\vskip 0.25cm
\begin{tabular}{l c c c c c c c}
             &   \multicolumn{7}{c}{$\Lambda$ (TeV)} \\
\cline{2-8}
Decay Mode   &  35  &  40  &  50  &  60 &  70  &  80  &  85  \\
\hline
$\chi_1^\pm \rightarrow \tilde{\tau}_1 \nu_\tau$  &  1  &  1  &  0.6787
             &  0.5192  &  0.4440  &  0.3996  &  0.3833  \\
$\chi_1^\pm \rightarrow \chi_1^0 \, W$  &  -  &  -  &  0.3213  &  0.4808
             &  0.5560  &  0.6004  &  0.6167  \\
\hline
$\chi_2^0 \rightarrow \tilde{\tau}_1 \tau$  &  0.5677  &  0.5965  &  0.6235
             &  0.3075  &  0.2137  &  0.1719  &  0.1578  \\
$\chi_2^0 \rightarrow \tilde{\mu}_1 \mu$  &  0.2162  &  0.2017  &  0.1660
             &  0.0659  &  0.0378  &  0.0256  &  0.0217  \\
$\chi_2^0 \rightarrow \tilde{e}_1 e$  &  0.2162  &  0.2017  &  0.1660
             &  0.0659  &  0.0378  &  0.0256  &  0.0217  \\
$\chi_2^0 \rightarrow \chi_1^0 \, Z$  &  -  &  -  &  0.0446  &  0.0318
             &  0.0251  &  0.0219  &  0.0207  \\
$\chi_2^0 \rightarrow \chi_1^0 \, {\rm h}$  &  -  &  -  &  -  &  0.5289
             &  0.6856 &  0.7550  &  0.7780  \\
\hline
$\chi_1^0 \rightarrow \tilde{\tau}_1 \tau$  &  1  &  1  &  0.8577  &  0.6542
             &  0.5659  &  0.5215  &  0.5072  \\
$\chi_1^0 \rightarrow \tilde{\mu}_1 \mu$  &  -  &  -  & 0.0711  &  0.1729
             &  0.2170  &  0.2392  &  0.2464  \\
$\chi_1^0 \rightarrow \tilde{e}_1 e$  &  -  &  -  & 0.0711  &  0.1729
             &  0.2170  &  0.2392  &  0.2464  \\
\hline
$\tilde{e}_1 \rightarrow \chi_1^0 \, e$   &  1  &  1  &  -  &  -  &  -  &  -
             &  -  \\
$\tilde{e}_1^- \rightarrow e^- \tau^- \tilde{\tau}^+$  &  -  &  -  &  0.5205
             &  0.5287  &  0.5315  & 0.5310  &  0.5298  \\
$\tilde{e}_1^- \rightarrow e^- \tau^+ \tilde{\tau}^-$  &  -  &  -  &  0.4795
             &  0.4697  &  0.4634  &  0.4580  &  0.4554  \\
$\tilde{e}_1 \rightarrow e \, \tilde{G}$  &  -  &  -  &  -  &  0.0016
             &  0.0050 &  0.0110  &  0.0148  \\
\end{tabular}
\end{table}

\begin{table}
\centering
\caption{\label{tau1br35} Inclusive tau-jet branching ratios for the
dominant production mechansims for the parameters $n = 2$, $\tan \beta = 15$,
$\Lambda = 35$\,TeV and $M = 105$\,TeV.}
\vskip 0.25cm
\begin{tabular}{l c c c c c}
Production Mode  &  1 $\tau$-jet  &  2 $\tau$-jets  &  3 $\tau$-jets
                 &  4 $\tau$-jets  &  5 $\tau$-jets  \\
\hline
$\chi_1^+ \chi_1^-$: no cuts  &  0.4562  &  0.4200 &  -  &  -
       &  -  \\
with cuts  &  0.2577  &  0.1084  &  -  &  -
       &  -  \\
\hline
$\chi_1^\pm \chi_2^0$: no cuts  &  0.2408  &  0.4434  &  0.2723
       &  -  &  -  \\
with cuts  &  0.2558  &  0.1567  &  0.0259
       &  -  &  -  \\
\hline
$\tilde{\tau}_1^+ \tilde{\tau}_1^-$: no cuts  &  0.4560  &  0.4203  &  -
       &  -  &  -  \\
with cuts    &  0.2523  &  0.0939  &  -  &  -  &  -  \\
\hline
$\tilde{e}_1^+ \tilde{e}_1^-$: no cuts  &  0.1128  &  0.3118  &  0.3834
      & 0.1766  &  -  \\
with cuts   &  0.2383  &  0.0778  &  0.0003  & {\it negl.}  &  -  \\
\end{tabular}
\end{table}

\begin{table}
\centering
\caption{\label{tau1br50} Inclusive tau-jet branching ratios for the
dominant production mechansims for the parameters $n = 2$, $\tan \beta = 15$,
$\Lambda = 50$\,TeV and $M = 150$\,TeV.}
\vskip 0.25cm
\begin{tabular}{l c c c c c}
Production Mode  &  1 $\tau$-jet  &  2 $\tau$-jets  &  3 $\tau$-jets
   &  4 $\tau$-jets  &  5 $\tau$-jets \\
\hline
$\chi_1^+ \chi_1^-$: no cuts  &  0.3194  &  0.4099  &  0.1627
     &  0.0291  &  0.0027  \\
with cuts  &  0.3355  &  0.1610  &  0.0100  &  0.0004  & {\it negl.} \\
\hline
$\chi_1^\pm \chi_2^0$: no cuts  &  0.1969  &  0.3961  &  0.3060
     &  0.0626  &  0.0044  \\
with cuts  &  0.3234  &  0.2238  &  0.0472  &  0.0014  & {\it negl.} \\
\hline
$\tilde{\tau}_1^+ \tilde{\tau}_1^-$: no cuts  &  0.4561  &  0.4201  &  -
     &  -  &  -  \\
with cuts  &  0.3345  &  0.1370  &  -  &  -  &  -  \\
\hline
$\tilde{e}_1^+ \tilde{e}_1^-$: no cuts  &  0.1130  &  0.3119
     &  0.3833  &  0.1765  &  -  \\
with cuts  &  0.3199  &  0.1207  &  0.0023  & {\it negl.} &  -  \\
\end{tabular}
\end{table}

\begin{table}
\centering
\caption{\label{tau1br70} Inclusive tau-jet branching ratios for the
dominant production mechansims for the parameters $n = 2$, $\tan \beta = 15$
$\Lambda = 70$\,TeV and $M = 210$\,TeV.}
\vskip 0.25cm
\begin{tabular}{l c c c c c}
Production Mode   &  1 $\tau$-jet  &  2 $\tau$-jets  &  3  $\tau$-jets
    &  4  $\tau$-jets  &  5  $\tau$-jets \\
\hline
$\chi_1^+ \chi_1^-$: no cuts  &  0.2333  &  0.3821  &  0.2549  &  0.0728
    &  0.0078  \\
with cuts    &  0.3647  &  0.2119  &  0.0322  &  0.0023  & {\it negl.} \\
\hline
$\chi_1^\pm \chi_2^0$: no cuts  &  0.1574  &  0.3455  &  0.3259
    &  0.1210  &  0.0202  \\
with cuts    &  0.3419  &  0.2334  &  0.0569  &  0.0071  &  0.0008  \\
\hline
$\tilde{\tau}_1^+ \tilde{\tau}_1^-$: no cuts  &  0.4561  &  0.4201  &  -
    &  -  &  -  \\
with cuts  &  0.3988  &  0.1711  &  -  &  -  &  -  \\
\hline
$\tilde{e}_1^+ \tilde{e}_1^-$: no cuts  &  0.1165  &  0.3133  &  0.3795
    &  0.1743  &  -  \\
with cuts  &  0.3901  &  0.1495  &  0.0013  & {\it negl.} &  -  \\
\end{tabular}
\end{table}

\begin{table}
\centering
\caption{\label{spec1br} Production rates in fb for some of the more
interesting final state configurations with and without cuts
for the parameters $n = 2$, $\tan \beta = 15$ and $M$/$\Lambda$ = 3.}
\vskip 0.25cm
\begin{tabular}{l c c c c}
    &  \multicolumn{2}{c}{$\Lambda = 35$\,TeV}
    &  \multicolumn{2}{c}{$\Lambda = 50$\,TeV}  \\
    &  no cuts  &  cuts  &  no cuts  &  cuts  \\
\hline
1 $\tau$-jet               & -      &  51.17   & -       & 3.71  \\
$e$/$\mu$ + 1 $\tau$-jet   & 85.44  &  11.91   & 4.98    & 1.59  \\
1 jet + 1 $\tau$-jet       & -      &  21.58   & -       & 2.36  \\
2 jets + 1 $\tau$-jet      & -      &  -       & -       & 2.35  \\
$e$/$\mu$+ 2 jets + 1 $\tau$-jet  &  -  &  -   & -       & 0.69  \\
2 $\tau$-jets              & 157.3  &  40.64   & 9.18    & 3.91   \\
\end{tabular}
\end{table}

\begin{table}
\centering
\caption{\label {br2} Branching ratios of some of the sparticles of
interest for the parameter set with $n = 3$, $\tan \beta = 15$ and
$M$/$\Lambda$ = 20.}
\vskip 0.25cm
\begin{tabular}{l c c c c c}
    &  \multicolumn{5}{c}{$\Lambda$ (TeV)}   \\
\cline{2-6}
Decay Mode     & 25 &  30 & 40 & 50 & 55  \\
\hline
$\chi_1^\pm \rightarrow \tilde{\tau}_1 \nu_\tau$  &  0.8062  &  0.6330  &
         0.3514  &  0.2211  &  0.1839  \\
$\chi_1^\pm \rightarrow \tilde{\tau}_2 \nu_\tau$  &  -  &  -  &  0.0118  &
         0.0493  &  0.0643  \\
$\chi_1^\pm \rightarrow \tilde{e}_2 \nu_e$        &  -  &  -  &  0.023  &
         0.061  &  0.074  \\
$\chi_1^\pm \rightarrow \tilde{\nu}_\tau \tau$  &  0.0729  &  0.1097  &
         0.1379  &  0.1450  &  0.1467 \\
$\chi_1^\pm \rightarrow \tilde{\nu}_e e$   &  0.0604  &  0.0954  &  0.1253 &
         0.1342  &  0.1365  \\
$\chi_1^\pm \rightarrow \chi_1^0 \, W$  &  -  &  0.0666  &  0.2022  &
         0.1943 & 0.1832 \\
\hline
$\chi_2^0 \rightarrow \tilde{\tau}_1 \tau$  &  0.5760  &  0.5653  & 0.2894 &
         0.1678  &  0.1385  \\
$\chi_2^0 \rightarrow \tilde{\tau}_2 \tau$  &  -  &  -  &  0.0142  &
         0.0432  &  0.0542  \\
$\chi_2^0 \rightarrow \tilde{e}_1 e$  &  0.1667  &  0.1335  &  0.0478  &
         0.0207  &  0.0151  \\
$\chi_2^0 \rightarrow \tilde{e}_2 e$  &  -  &  -  &  0.0240  &  0.0507  &
         0.0602  \\
$\chi_2^0 \rightarrow \tilde{\nu}_\tau \nu_\tau$  &  0.0317  &  0.0583  &
         0.0760  &  0.0811  &  0.0845 \\
$\chi_2^0 \rightarrow \tilde{\nu}_e \nu_e$  &  0.0294  &  0.0547  &
         0.0722  &  0.07764  &  0.0810 \\
$\chi_2^0 \rightarrow \chi_1^0 \, Z$  &  -  &  -  &  0.0196  &  0.0136  &
         0.0119  \\
$\chi_2^0 \rightarrow \chi_1^0 \, {\rm h}$  &  -  &  -  &  0.3127  &  0.3961  &
         0.3983  \\
\hline
$\chi_1^0 \rightarrow \tilde{\tau}_1 \tau$  &  0.7955  &  0.6011  &
         0.4707  &  0.4268  &  0.4150  \\
$\chi_1^0 \rightarrow \tilde{e}_1 e$  &  0.1023  &  0.1994  &  0.2646  &
         0.2866  &  0.2925  \\
\hline
$\tilde{e}^- \rightarrow e^- \tau^- \tilde{\tau}^+$  &  0.5563  &  0.5746  &
        0.5898  &  0.5960  &  0.5977  \\
$\tilde{e}^- \rightarrow e^- \tau^+ \tilde{\tau}^-$  &  0.4437  &  0.4254  &
        0.4102  &  0.4040  &  0.4023  \\
\end{tabular}
\end{table}

\begin{table}
\centering \caption{\label{tau2br25} Inclusive tau-jet branching
ratios for the dominant production mechansims for the parameters
$n = 3$, $\tan \beta = 15$, $\Lambda = 25$\,TeV and $M =
500$\,TeV.} \vskip 0.25cm
\begin{tabular}{l c c c c c}
Production Mode   &   1 $\tau$-jet  &  2  $\tau$-jets  &  3  $\tau$-jets
    &   4  $\tau$-jets  &  5  $\tau$-jets  \\
\hline
$\chi_1^+ \chi_1^-$: no cuts  &  0.3597  &  0.4040  &  0.1110  &  0.0306
    &  0.0033  \\
with cuts  &  0.2336  &  0.1138  &  0.0077  &  0.0003  & {\it negl.} \\
\hline
$\chi_1^\pm \chi_2^0$: no cuts  &  0.2110  &  0.4077  &  0.2912
    &  0.0443  &  0.0085  \\
with cuts  &  0.2452  &  0.1682  &  0.0332  &  0.0010  & {\it negl.} \\
\hline
$\tilde{\tau}_1^+ \tilde{\tau}_1^-$: no cuts  &  0.4555  &  0.4210  &  -
    &  -  &  -  \\
with cuts  &  0.2258  &  0.0797  &  -  &  -  &  -  \\
\hline
$\tilde{e}_1^+ \tilde{e}_1^-$: no cuts  &  0.1128  &  0.3121  &  0.3833
    &  0.1764  &  -  \\
with cuts  &  0.1998  &  0.0779  &  0.0063  &  0.0002  &  -  \\
\end{tabular}
\end{table}

\begin{table}
\centering
\caption{\label{tau2br40} Inclusive tau-jet branching ratios for the
dominant production mechansims for the parameters $n = 3$, $\tan \beta = 15$
$\Lambda = 40$\,TeV and $M = 800$\,TeV.}
\vskip 0.25cm
\begin{tabular}{l c c c c c}
Production Mode    &  1  $\tau$-jet  &  2  $\tau$-jet  &  3  $\tau$-jets
     &  4  $\tau$-jets  &  5  $\tau$-jets  \\
\hline
$\chi_1^+ \chi_1^-$: no cuts  &  0.1890  &  0.3484  &  0.2863
    &  0.1155  &  0.0223  \\
with cuts    &  0.3011  &  0.2123  &  0.0500  &  0.0057  &  0.0003  \\
\hline
$\chi_1^\pm \chi_2^0$: no cuts  &  0.1457  &  0.3342  &  0.3301
    &  0.1395  &  0.0236  \\
with cuts    &  0.2991  &  0.2343  &  0.0672  &  0.0082  &  0.0008  \\
\hline
$\tilde{\tau}_1^+ \tilde{\tau}_1^-$: no cuts  &  0.4556  &  0.4208
    &  -  &  -  &  -  \\
with cuts    &  0.3396  &  0.1395  &  -  &  -  &  -  \\
\hline
$\tilde{e}_1^+ \tilde{e}_1^-$: no cuts  &  0.1128  &  0.3121  &  0.3833
    &  0.1765  &  -  \\
with cuts    &  0.3213  &  0.1270  &  0.0048  &  0.0001  &  -  \\
\end{tabular}
\end{table}

\begin{table}
\centering
\caption{\label{tau3br25}  Inclusive $\tau$-jet branching ratios for
the various production mechanisms
for the parameters $n = 3$, $\tan \beta = 3$, $\Lambda = 25$\,TeV
and $M = 75$\,TeV.}
\vskip 0.25cm
\begin{tabular}{l c c c c c}
Production Mode &  1  $\tau$-jet  &  2  $\tau$-jets  &  3  $\tau$-jets
      &  4  $\tau$-jets  &  5  $\tau$-jets  \\
\hline
$\chi_1^+ \chi_1^-$: no cuts  &  0.4562  & 0.4200  &  -  &  -  &  -  \\
with cuts    &  0.2514  &  0.1053  &  -  &  -  &  -  \\
\hline
$\chi_1^\pm \chi_2^0$: no cuts  &  0.5088  &  0.1514  &  0.0935
      &  -  &  -  \\
with cuts     &  0.2480  &  0.0666  &  0.0153  &  -  &  -  \\
\hline
$\tilde{\tau}_1^+ \tilde{\tau}_1^-$: no cuts  &  0.4560  &  0.4203
      &  -  &  -  &  -  \\
with cuts    &  0.2427  &  0.0891  &  -  &  -  &  -  \\
\end{tabular}
\end{table}

\begin{table}
\centering
\caption{\label{tau3br40}  Inclusive $\tau$-jet branching ratios for
the various production mechanisms
for the parameters $n = 3$, $\tan \beta = 3$, $\Lambda = 40$\,TeV
and $M = 120$\,TeV.}
\vskip 0.25cm
\begin{tabular}{l c c c c c}
Production Mode &  1  $\tau$-jet  &  2  $\tau$-jets  &  3  $\tau$-jets
        &  4  $\tau$-jets  &  5  $\tau$-jets  \\
\hline
$\chi_1^+ \chi_1^-$: no cuts  &  0.2903  &  0.2272  &  0.0937
      &  0.0299  &  0.0054  \\
with cuts    &  0.1826  &  0.0996  &  0.0227  &  0.0034  &  0.0002  \\
\hline
$\chi_1^\pm \chi_2^0$: no cuts  &  0.2613  &  0.2299  &  0.0826
        &  0.0298  &  0.0057  \\
with cuts    &  0.1903  &  0.1054  &  0.0232  &  0.0038  &  0.0004  \\
\hline
$\tilde{\tau}_1^+ \tilde{\tau}_1^-$: no cuts  &  0.4560  &  0.4202
        &  -  &  -  &  -  \\
with cuts    &  0.3419  &  0.1410  &  -  &  -  &  -  \\
\end{tabular}
\end{table}

\begin{table}
\centering
\caption{\label{co_br} Branching ratios of some of the sparticles of
interest for the parameter set with $n = 3$, $\tan \beta = 3$ and
$M$/$\Lambda$ = 3.}
\vskip 0.25cm
\begin{tabular}{l c c c c c c}
    &  \multicolumn{6}{c}{$\Lambda$ (TeV)}  \\
\cline{2-7}
Decay Mode     & 25 & 30 & 40 & 50 & 60 & 65 \\
\hline
$\chi_1^\pm \rightarrow \tilde{\tau}_1 \nu_\tau$  &  1  &  0.3416
               &  0.0225  &  0.0099  &  0.0063  &  0.0053  \\
$\chi_1^\pm \rightarrow \tilde{\nu}_\tau \tau$    &  -  &  0.2147
               &  0.1426  &  0.1390  &  0.1402  &  0.1410  \\
$\chi_1^\pm \rightarrow \tilde{\nu}_e e$    &  -  &  0.2219
               &  0.1418  &  0.1383  &  0.1396  &  0.1405  \\
$\chi_1^\pm \rightarrow \chi_1^0 \, W$    &  -  &  -  &  0.4514
               &  0.3436  &  0.2781  &  0.2552  \\
$\chi_1^\pm \rightarrow \tilde{\tau}_2 \nu_\tau$  &  -  &  -
               &  0.0326  &  0.0765  &  0.0984  &  0.1055  \\
$\chi_1^\pm \rightarrow \tilde{e}_2 e$  &  -  &  -  &  0.0336
               &  0.0772  &  0.0989  &  0.1060  \\
\hline
$\chi_2^0 \rightarrow \tilde{\tau}_1 \tau$  &  0.3340  &  0.3135
               &  0.1384  &  0.0577  &  0.0302  &  0.0232  \\
$\chi_2^0 \rightarrow \tilde{e}_1 e$  &  0.3223  &  0.2977  &
               0.1265  &  0.0505  &  0.0252  &  0.0190  \\
$\chi_2^0 \rightarrow \tilde{\nu}_\tau \nu_\tau$  &  0.0071  &  0.0304
               &  0.0820  &  0.0987  &  0.1062  &  0.1086  \\
$\chi_2^0 \rightarrow \tilde{\nu}_e \nu_e$  &  0.0071  &  0.0303
               &  0.0819  &  0.0986  &  0.1061  &  0.1085  \\
$\chi_2^0 \rightarrow \chi_1^0 \, Z$  &  -  &  -  &  0.0124  &  0.0083
               &  0.0060  &  0.0053  \\
$\chi_2^0 \rightarrow \chi_1^0 \, {\rm h}$  &  -  &  -  &  0.2146  &  0.2763
               &  0.2800  &  0.2779  \\
$\chi_2^0 \rightarrow \tilde{\tau}_2 \tau$  &  -  &  -  &  0.0447
               &  0.0866  &  0.1049  &  0.1100  \\
$\chi_2^0 \rightarrow \tilde{e}_2 e$  &  -  &  -  &  0.0456  &  0.0871
               &  0.1051  &  0.1101  \\
\hline
$\chi_1^0 \rightarrow \tilde{\tau}_1 \tau$  &  0.3593  &  0.3437
               &  0.3378  &  0.3362  &  0.3355  &  0.3353  \\
$\chi_1^0 \rightarrow \tilde{e}_1 e$  &  0.3203  &  0.3281  &  0.3311
               &  0.3319  &  0.3322  &  0.3323  \\
\hline
$\tilde{\nu}_\tau \rightarrow \chi_1^\pm \tau$  &  0.0047  &  -  &  -
               &  -  &  -  &  -  \\
$\tilde{\nu}_\tau \rightarrow \nu_\tau \, \chi_1^0$  &  0.9953  &  0.9940
               &  0.9889  &  0.9875  &  0.9870  &  0.9869  \\
$\tilde{\nu}_\tau \rightarrow \tilde{\tau}_1 W$  &  -  &  0.0060
               &  0.0111  &  0.0125  &  0.0130  &  0.0131  \\
\hline
$\tilde{\nu}_e \rightarrow \chi_1^0 \nu_e$  &  0.9928  &  1  &  1
               &  1  &  1  &  1  \\
$\tilde{\nu}_e \rightarrow \chi_1^\pm e$  &  0.0072  &  -  &  -  &  -
               &  -  &  -  \\
\hline
$\tilde{\tau}_2 \rightarrow \chi_1^0 \tau$  &  0.4640  &  0.8960
               &  1  &  1  &  1  &  1  \\
$\tilde{\tau}_2 \rightarrow \chi_1^\pm \nu_\tau$  &  0.4155  &  0.0983
               &  -  &  -  &  -  &  -  \\
$\tilde{\tau}_2 \rightarrow \chi_2^0 \tau$  &  0.1206  &  0.0057
               &  -  &  -  &  -  &  -  \\
\hline
$\tilde{e}_2 \rightarrow \chi_1^0 e$  &  1  &  1  &  1  &  1  &  1  &  1  \\
\hline
$\tilde{e}_1 \rightarrow \tilde{G} e$  &  1  &  1  &  1  &  1  &  1  &  1  \\
\end{tabular}
\end{table}

\begin{table}
\centering
\caption{\label{spec3br} Production rates in fb for some of the more
interesting final state configurations with and without cuts
for the parameters $n = 3$, $\tan \beta = 3$ and $M$/$\Lambda$ = 3.}
\vskip 0.25cm
\begin{tabular}{l c c c c}
     &   \multicolumn{2}{c}{$\Lambda = 25$\,TeV}
     &   \multicolumn{2}{c}{$\Lambda = 30$\,TeV}  \\
     &  no cuts  &  cuts  &  no cuts  &  cuts  \\
\hline
$\tau$-jet      &  -      &  28.84  &  -     &  5.39  \\
2 $\tau$-jets   &  70.57  &  23.11  &  8.01  &  4.49  \\
$e$/$\mu$ \& $\tau$-jet   &  38.20  &  12.71 &  4.35  &  3.82  \\
2$e$/2$\mu$ \& $\tau$-jet & -   &  6.48  &  3.99  &  2.75  \\
\end{tabular}
\end{table}


\begin{figure}
\centering
\epsfxsize=0.6\textwidth
\epsfbox[84 233 487 544]{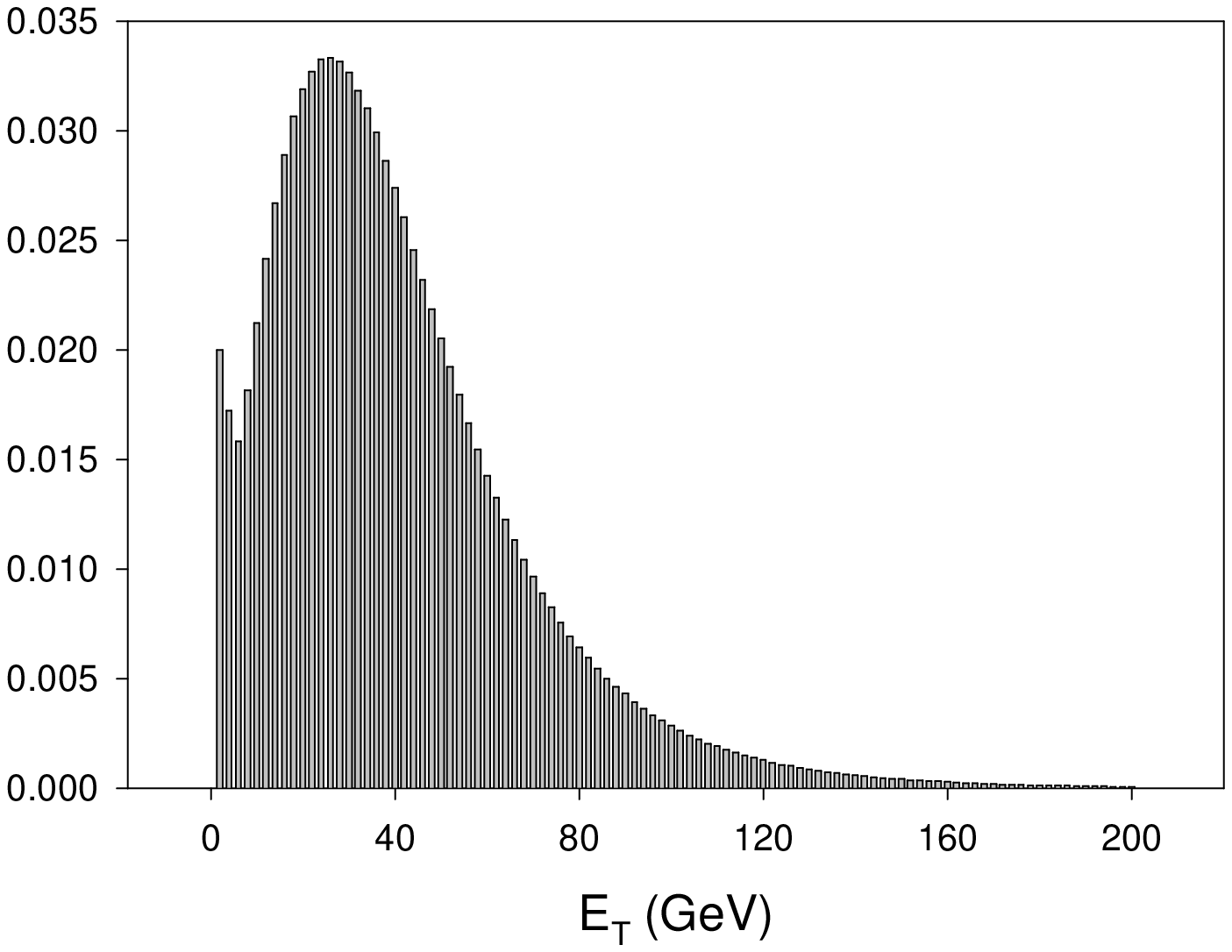}
\begin{minipage}{0.6\textwidth}
(a)
\end{minipage}
\end{figure}

\vspace{0.5 cm}

\begin{figure}
\centering
\epsfxsize=0.6\textwidth
\epsfbox[89 233 487 544]{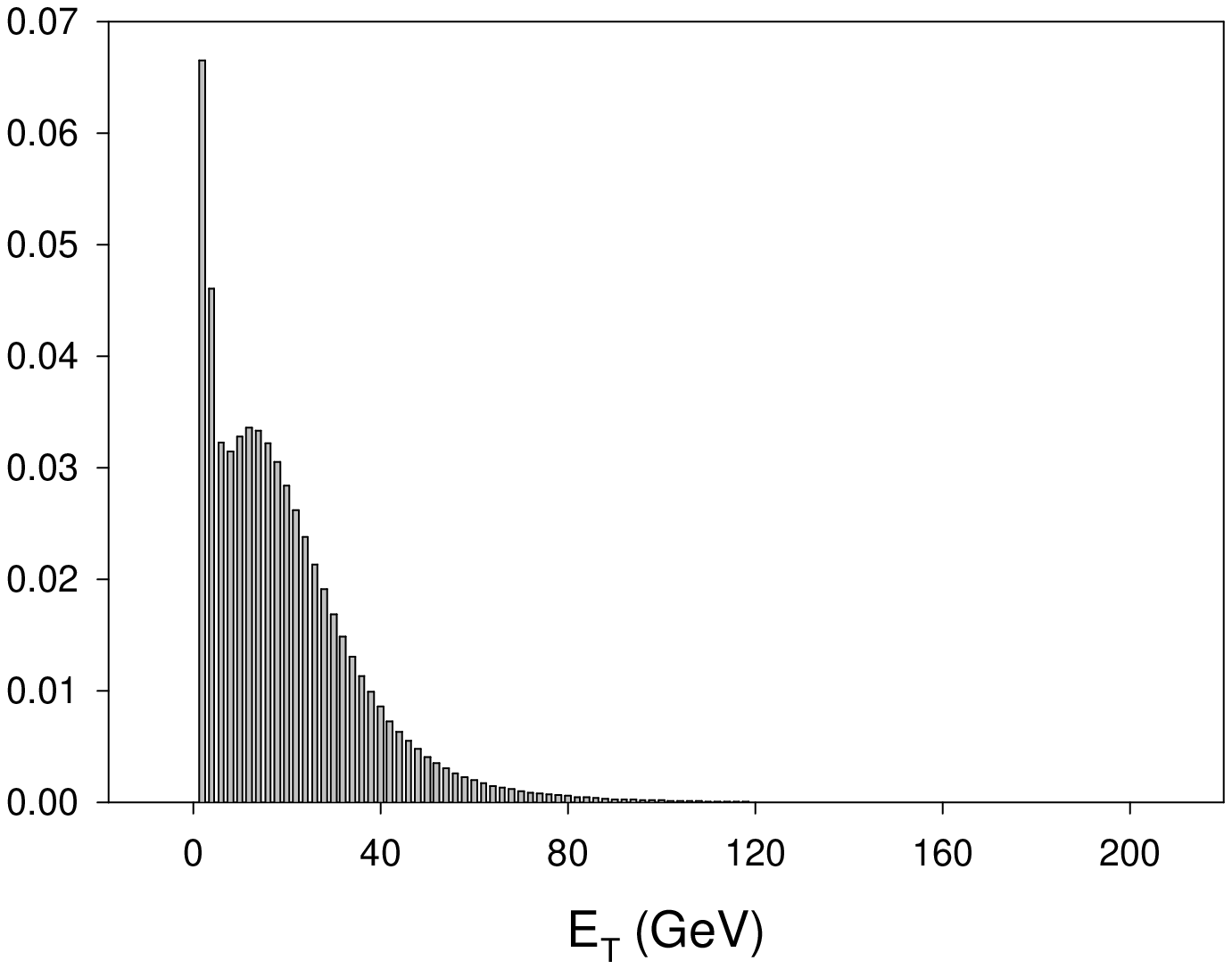}
\begin{minipage}{0.6\textwidth}
(a)
\end{minipage}
\end{figure}

\vspace{0.25 cm}

\begin{figure}
\centering
\epsfxsize=0.6\textwidth
\epsfbox[89 233 487 544]{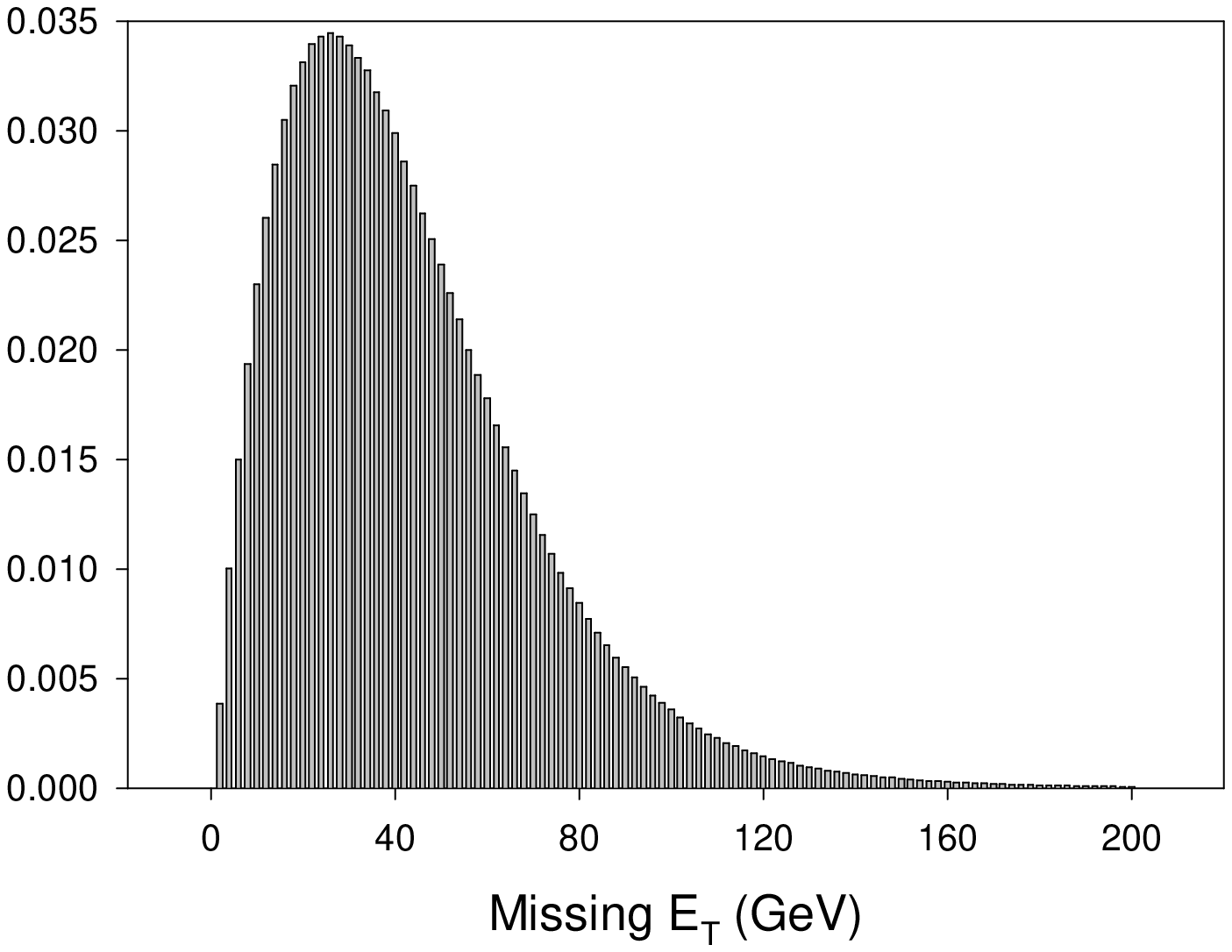}
\begin{minipage}{0.6\textwidth}
(a)
\end{minipage}
\end{figure}

\vspace{0.5 cm}

\clearpage

\begin{figure}
\centering
\epsfxsize=0.98\textwidth
\epsfbox[67 134 500 646]{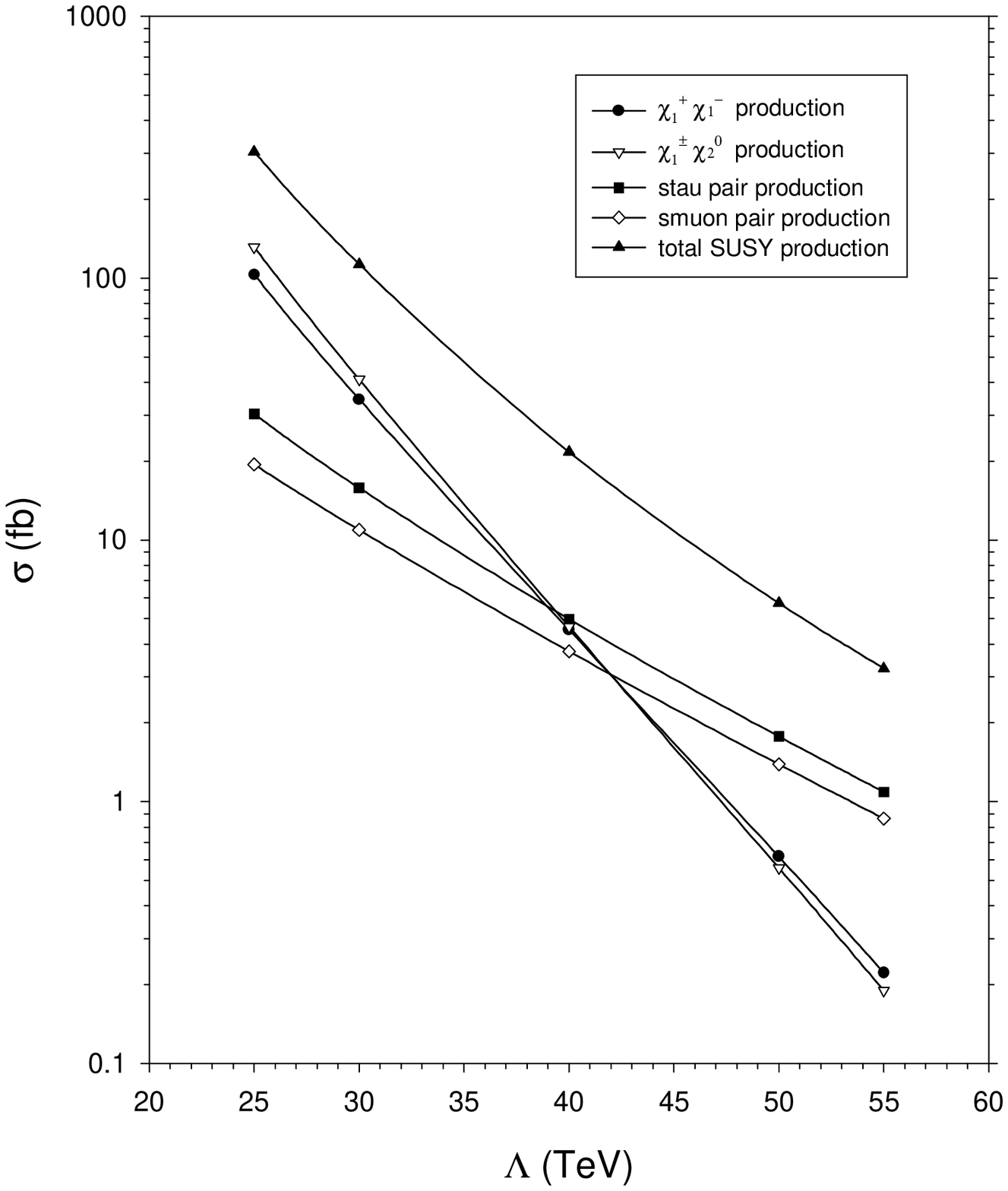}
\vskip 0.25cm
\caption{\label{cross2} Cross section for the important SUSY production
processes at the Tevatron for the line defined by $n = 3$,
$\tan \beta = 15$ and $M$/$\Lambda$ = 20. The $\chi_2^0 \chi_1^\pm$ cross
section includes production of both signs of the chargino.}
\end{figure}

\begin{figure}
\centering
\epsfxsize=0.6\textwidth
\epsfbox[84 233 488 544]{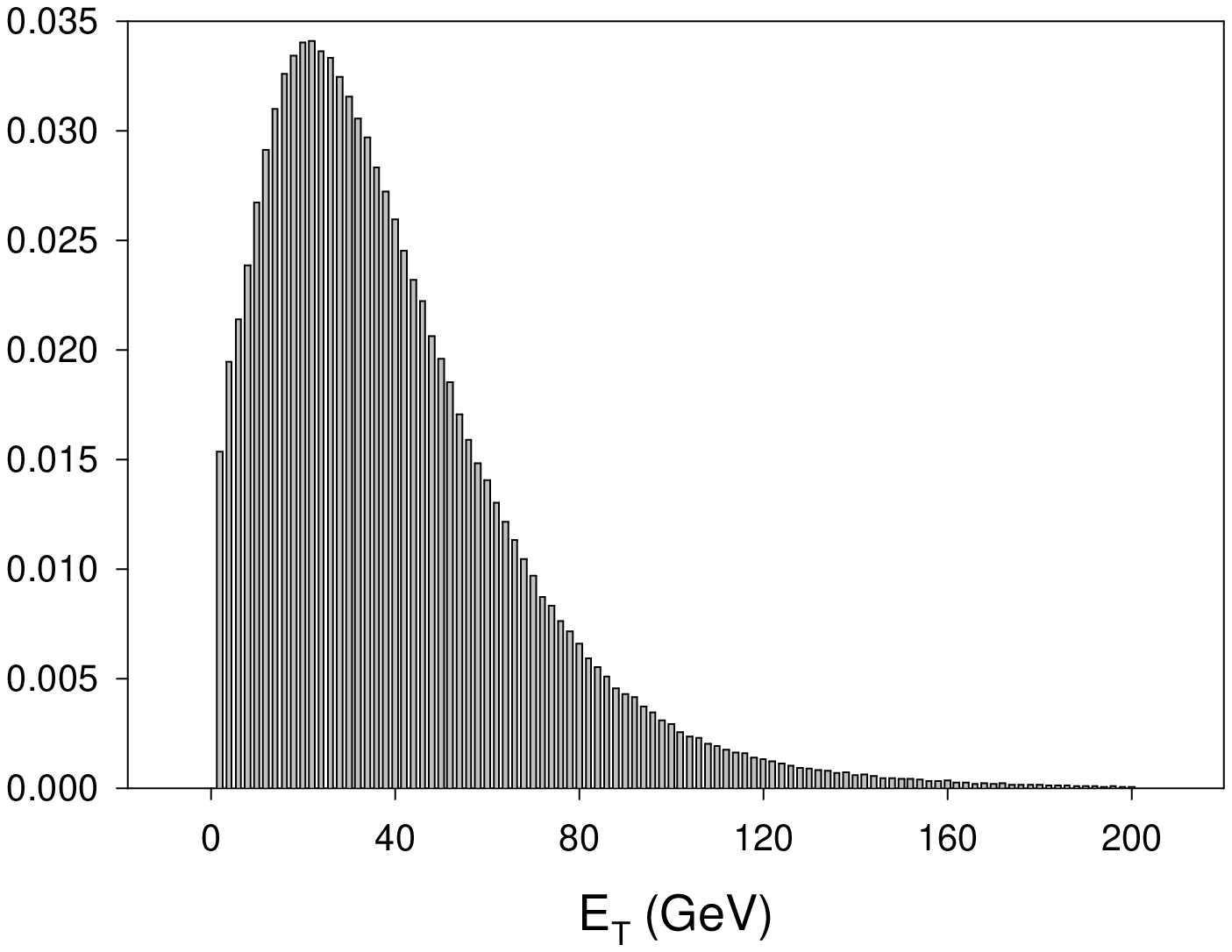}
\begin{minipage}{0.6\textwidth}
(a)
\end{minipage}
\end{figure}

\vspace{0.5 cm}

\begin{figure}
\centering
\epsfxsize=0.6\textwidth
\epsfbox[89 232 497 545]{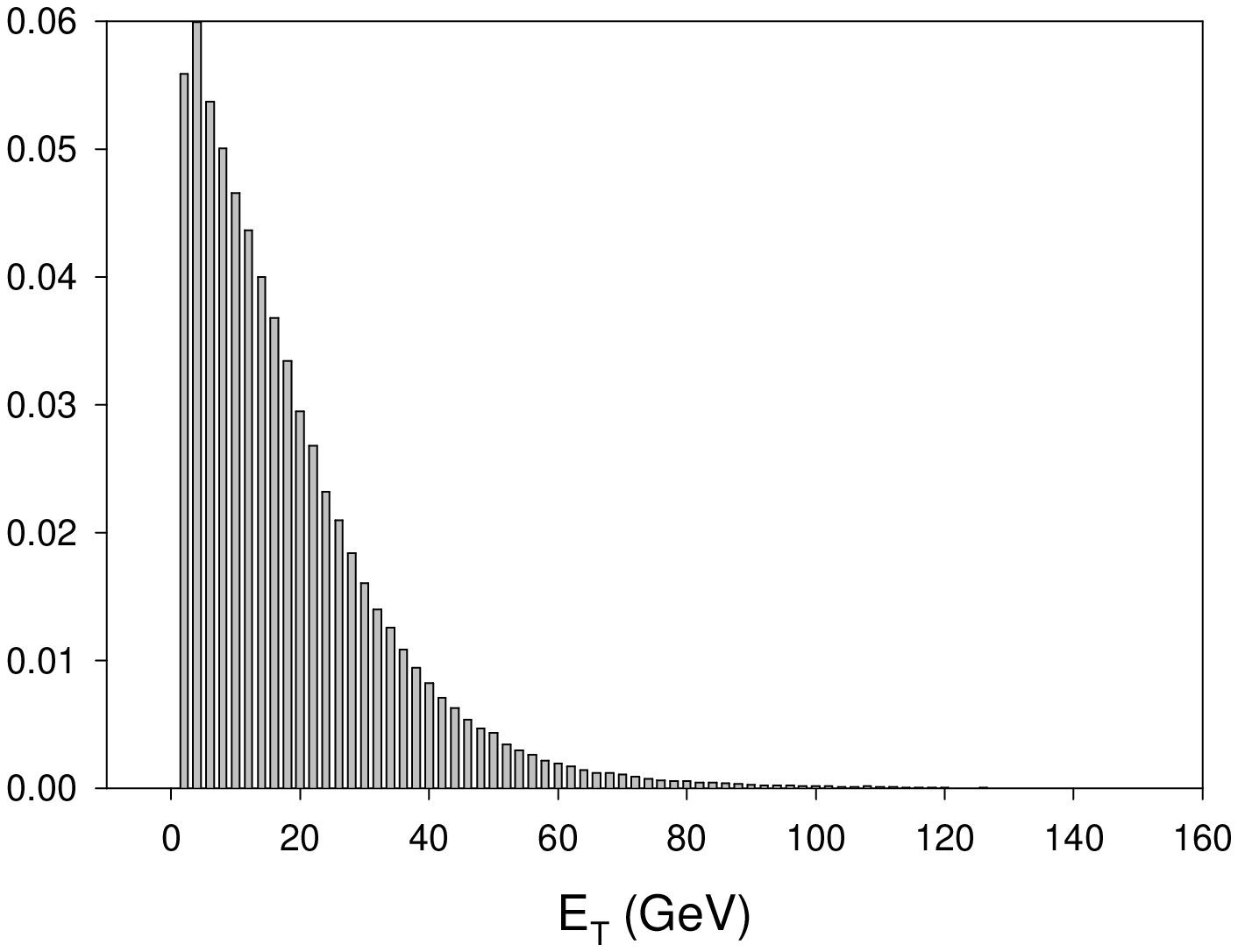}
\begin{minipage}{0.6\textwidth}
(a)
\end{minipage}
\end{figure}

\vspace{0.5 cm}

\begin{figure}
\centering
\epsfxsize=0.6\textwidth
\epsfbox[82 233 490 544]{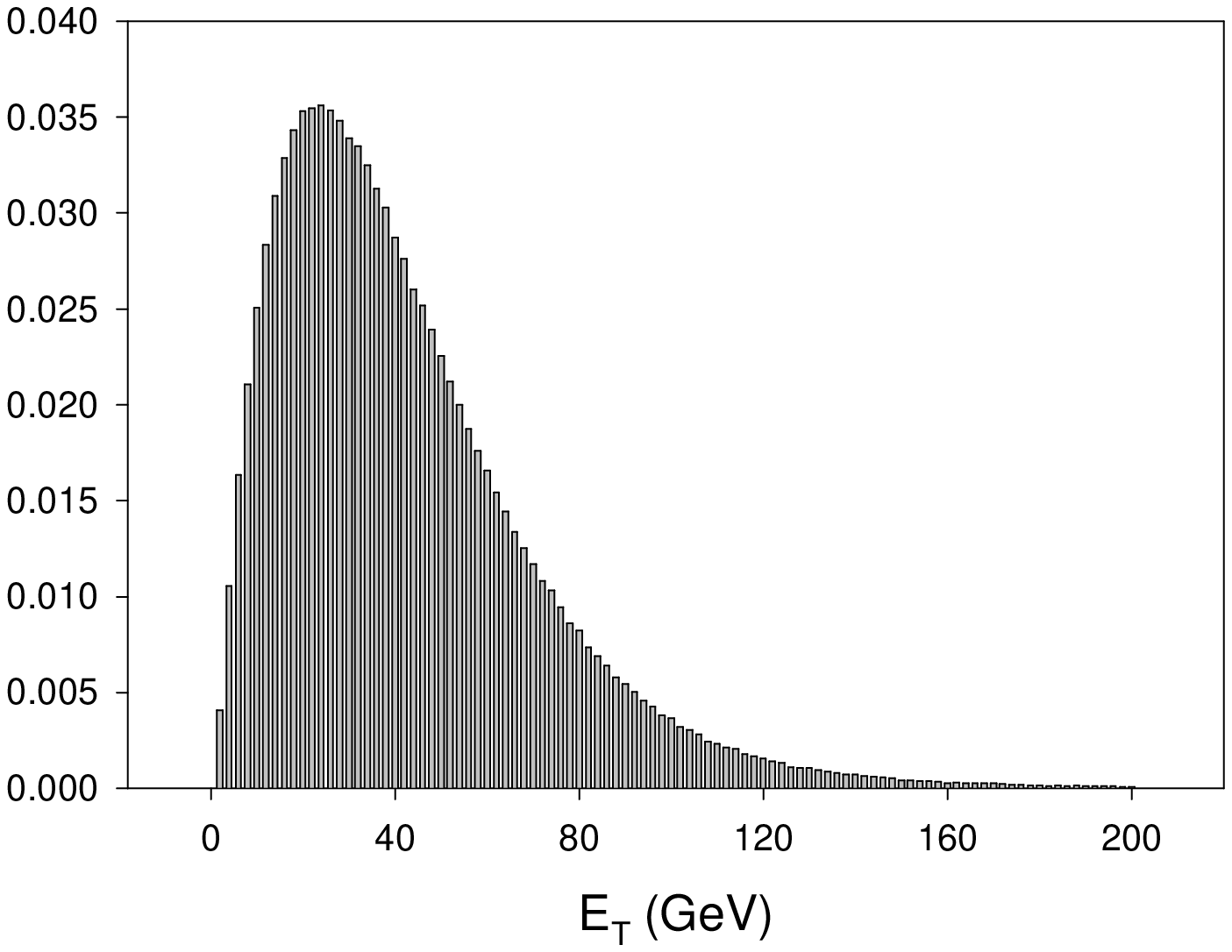}
\begin{minipage}{0.6\textwidth}
(a)
\end{minipage}
\end{figure}

\vspace{0.5 cm}

\clearpage

\begin{figure}
\centering
\epsfxsize=0.6\textwidth
\epsfbox[84 233 487 544]{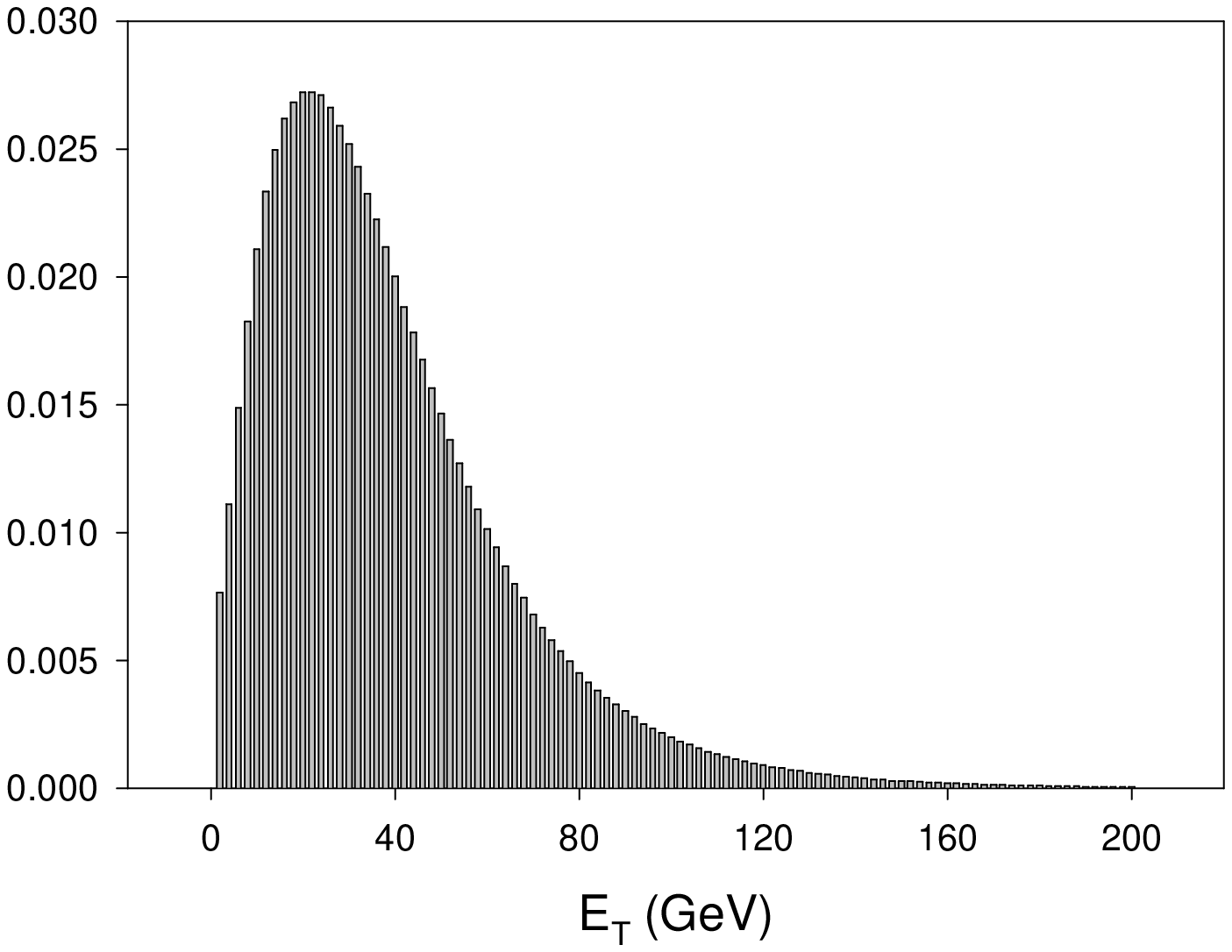}
\begin{minipage}{0.6\textwidth}
(a)
\end{minipage}
\end{figure}

\vspace{0.5 cm}

\begin{figure}
\centering
\epsfxsize=0.6\textwidth
\epsfbox[82 232 488 545]{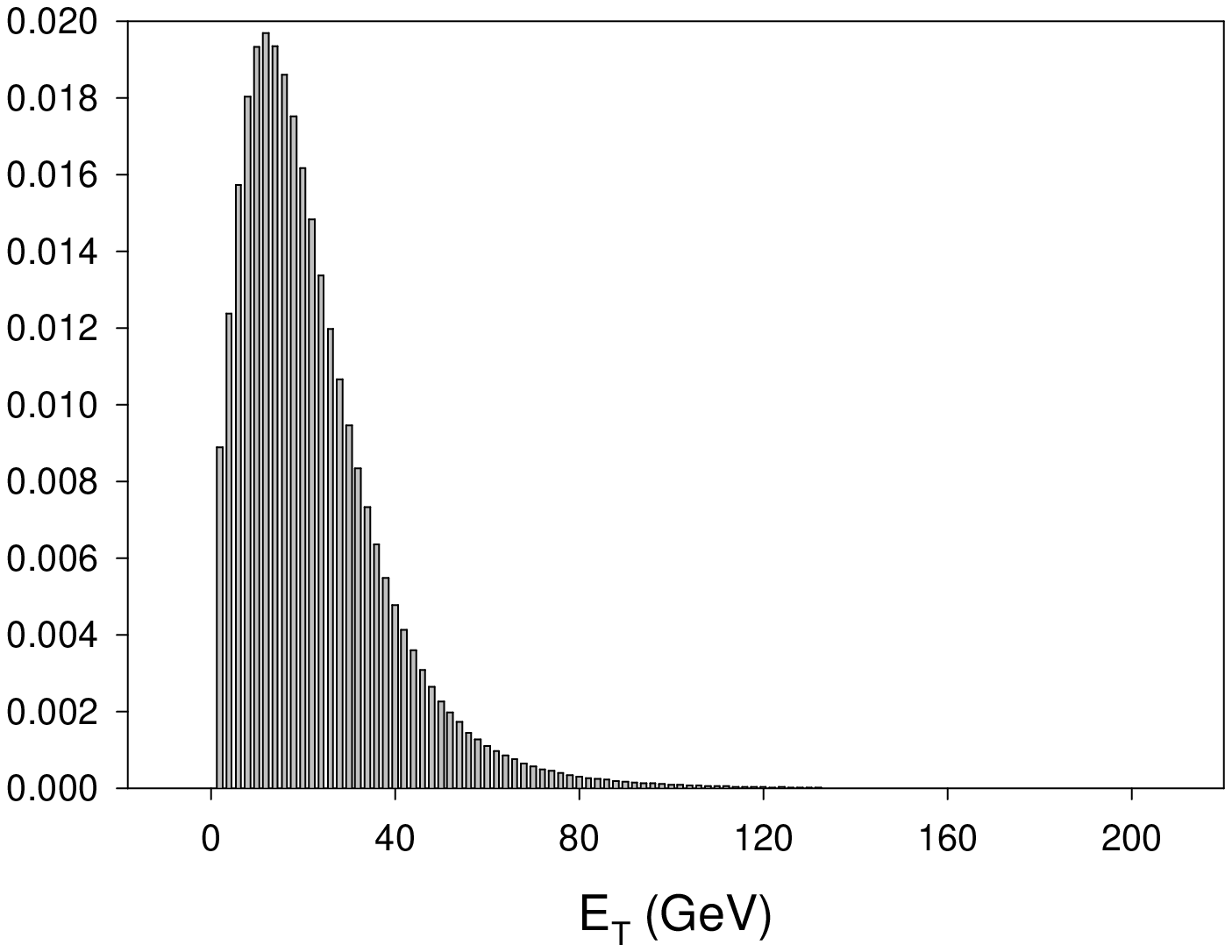}
\begin{minipage}{0.6\textwidth}
(a)
\end{minipage}
\end{figure}

\vspace{0.5 cm}

\begin{figure}
\centering
\epsfxsize=0.6\textwidth
\epsfbox[82 233 490 544]{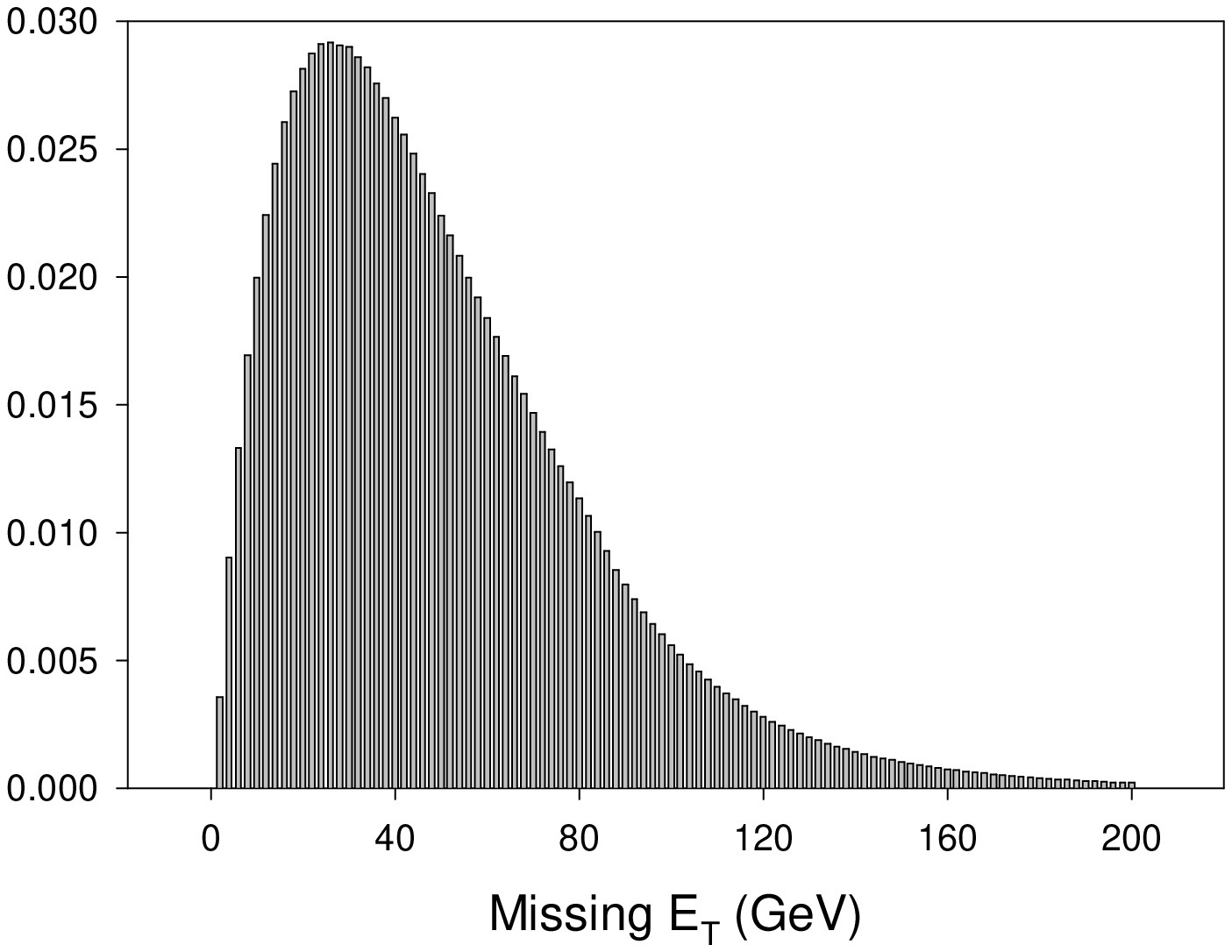}
\begin{minipage}{0.6\textwidth}
(a)
\end{minipage}
\end{figure}

\vspace{0.5 cm}

\end{document}